\documentclass[showpacs,amsmath,amssymb,aps,pra,10pt,reprint,superscriptaddress]{revtex4-1}
\usepackage{graphicx}
\usepackage{bm}
\usepackage[breaklinks=true,colorlinks=true,linkcolor=blue,urlcolor=blue,citecolor=blue]{hyperref}
\usepackage{graphics}
\usepackage{dcolumn}
\usepackage{amsmath,amssymb}
\usepackage{mathdots}
\usepackage{natbib}
\usepackage{soul,color}
\usepackage{epstopdf}
\usepackage[note-name]{notes2bib}
\usepackage{mathptmx} 

\begin{document}

\title{Nonreciprocal magnon fluxonics upon ferromagnet/superconductor hybrids}

\author{Oleksandr V. Dobrovolskiy}
\author{Andrii V. Chumak}
\affiliation{Faculty of Physics, University of Vienna, Boltzmanngasse 5, 1090 Vienna, Austria}

\begin{abstract}
Ferromagnet/superconductor heterostructures allow for the combination of unique physical phenomena offered by the both fields of magnetism and superconductivity. It was shown recently that spin waves can be efficiently scattered in such structures by a lattice of static or moving magnetic flux quanta (Abrikosov vortices), resulting in bandgaps in the spin-wave spectra. Here, we realize a nonreciprocal motion of a vortex lattice in nanoengineered symmetric and asymmetric pinning landscapes and investigate the non-reciprocal scattering of magnons on fluxons. We demonstrate that the magnon bandgap frequencies can be tuned by the application of a low-dissipative transport current and by its polarity reversal. Furthermore, we exploit the rectifying (vortex diode or ratchet) effect by the application of a 100\,MHz-frequency ac current to deliberately realize bandgap up- or downshifts during one ac halfwave while keeping the bandgap frequency constant during the other ac halfwave. The investigated phenomena allow for the realization of energy-efficient hybrid magnonic devices, such as microwave filters with an ultra-high bandgap tunability of 10\,GHz/mA and a fast modulation of the transmission characteristics on the $10$\,ns time scale.
\end{abstract}
\maketitle

\section{Introduction}
Spin waves, and their quanta magnons, are of great interest as potential data carriers in future low-energy data processing devices \cite{Bar21pcm,Mah20jap}. The phase of a spin wave and its pronounced nonlinear properties provide additional degrees of freedom \cite{Wan20nel,Dob19ami,Pap20arx,Wan21nac}, while the scalability of 2D and 3D structures \cite{Dob21apl,Hei20nal} and wavelengths \cite{Che20nac,Bau21apl,Liu18nac,Dob21arx} down to the nanometer regime are further advantages. Moreover, the utilization of macroscopic quantum states like Bose-Einstein condensation of magnons \cite{Dem06nat,Sch20nan,Sch21arx} and quantum operations with single magnons \cite{Lac20sci} are very promising. In this context, low-energy manipulation of spin waves attracts great attention. In particular, the manipulation of spin waves by electric fields is nowadays a subject of extensive investigations \cite{Ran19prb,Che17nal,Mer21acs}. A complementary low-energy approach is based on the use of electric currents in superconducting structures with vanishingly small resistance \cite{Gol18afm,Dob19nph}. Recently, it was demonstrated that spin waves can be efficiently manipulated by a lattice of Abrikosov vortices, so-called fluxons \cite{Bra95rpp,Abr04rmp}, in a ferromagnet/superconductor (F/S) Py/Nb heterostructure \cite{Dob19nph}. In this system, the magnon frequency spectrum exhibits a Bloch-like band structure \cite{Chu17jpd} that can be tuned by the biasing magnetic field. Furthermore, Bragg scattering of spin waves on a current-driven vortex lattice is accompanied by Doppler shifts of the band structure. This research direction is now called magnon fluxonics, referring to the data processing using the interaction between fluxons and magnons. The utilization of ultra-fast vortices \cite{Dob20nac} in magnon fluxonics has allowed for the recent observation of the Cherenkov generation of spin waves in F/S heterostructures \cite{Dob21arx}.

One of the primary advantages offered by spin waves for operations with data is a rich palette of nonreciprocal phenomena, implying that the spin-wave transport in opposite directions has different properties. For instance, nonreciprocity is inherent to the classical dipolar Damon-Eschbach spin-wave mode \cite{Esh60prv,Moh20prb}, to nano-structures with complex waveguide cross-sections \cite{Ota16prl,Hei21apl}, magnetic bilayers \cite{Gla16prb}, to systems with pronounced Dzyaloshinkski-Moria interactions \cite{Dik15apl,Bot21tom} etc. In addition, scattering of spin waves by a moving object also introduces a non-reciprocity associated with the co- or counter-propagation of the wave with respect to the moving object and with the sign of the resulting frequency Doppler shift \cite{Sta06prb,Chu10prb,Vla08sci}. In this regard, fluxonics \cite{Dob17pcs} offers unique opportunities, since vortices in nanoengineered superconductors represent a valuable playground for investigations of rectified net transport in ac-driven systems lacking reflection symmetry -- vortex ratchets \cite{Plo09tas,Shk14pcm,Dob15met,Dob20pra}. In these, the difference in the current values required to put vortices into motion against steep and gentle slopes of the pinning potential leads to the appearance of a rectifying voltage in response to an ac current drive (vortex diode effect). So far, vortex ratchet effects have extensively been investigated and their rectifying properties extend to the lower GHz ac frequency range \cite{Wor12prb,Dob15apl}.

Here, we combine a nonreciprocal spin-wave scattering on a lattice of moving vortices in Py/Nb heterostructures with a nonreciprocal motion of the vortex lattice in an asymmetric washboard pinning potential to realize a hybrid system with highly-tunable spin-wave spectra. We demonstrate that the application and polarity reversal of a dc current of 100\,$\mu$A results in a bandgap shift of about 2\,GHz. In the vortex ratchet regime, the application of a 100\,MHz-frequency ac current with an amplitude of 23\,$\mu$A allows for a fast modulation of the transmission characteristics on the $10$\,ns time scale. This enables the realization of bandgap up- or downshifts by about 0.1\,GHz during one ac halfwave while keeping the bandgap frequency constant during the other ac halfwave.

\section{Results and discussion}

\begin{figure}[t!]
    \centering
    \includegraphics[width=0.95\linewidth]{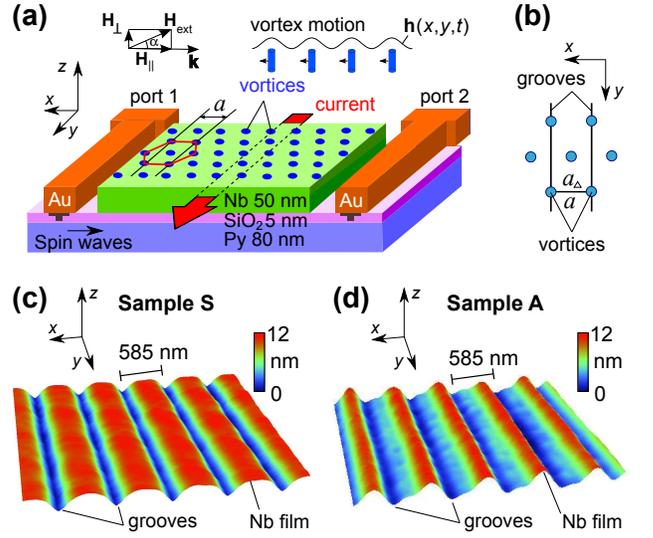}
    \caption{(a) Sketch of the investigated ferromagnet/superconductor Py/Nb heterostructure. The bilayer is in an inclined magnetic field $\mathbf{H}_\mathrm{ext}$ with in-plane component $H_{||}= 59.5$\,mT and out-of-plane component $H_\perp = 7.2$\,mT. Backward volume magnetostatic spin waves are excited by antenna 1, propagate through the Py waveguide, and are detected by antenna 2. The vortex lattice induces a spatially periodic magnetic field $\mathbf{h}(x,y)$ in Py, which becomes alternating in time when the vortices move under the action of the transport current.
    (b) Vortex lattice configuration at the matching field $H_\perp = 7.2$\,mT.
    (c) and (d) Atomic force microscope images of the symmetric (S) and asymmetric (A) washboard pinning nanolandscapes milled by focused ion beam on the surface of the Nb microstrips.}
    \label{f1}
\end{figure}

We investigate the coupled dynamics of spin waves and magnetic flux quanta in F/S bilayer structures consisting of a 80-nm-thick layer of ferromagnetic Py (F) and a 50-nm-thick superconducting Nb film (S), Fig.\,\ref{f1}(a). The Py layer acts as host for spin waves while the Nb layer harbors an Abrikosov vortex lattice whose presence and motion affects the spin-wave propagation. The F and S layers are coupled via stray fields, being electrically insulated from each other to avoid proximity effects \cite{Buz05rmp,Kom14apl}.

Superconducting vortices offer a unique tunable laboratory for studying the effects of different vortex lattice regimes on the propagation of spin waves. For an efficient control of magnetic flux quanta, we fabricated periodic arrays of nanogrooves on the surface of the S layers. In these nanolandscapes, the vortices are pinned at the groove bottoms because of the combined effect of vortex length reduction and suppressed superconducting order parameter owing to the implantation of Ga ions\,\cite{Dob12njp}. When a transport current is applied to the S layer, the vortex lattice can be put into motion. Specifically, a transport current $I$ applied along the $y$-axis in a magnetic field $\mathbf{H}_\mathrm{\perp}\equiv \mathbf{H}_z$ exerts on a vortex a Lorentz-type force (per unit length) $\mathbf{F}_\mathrm{L} = \Phi_0 [\mathbf{j}\times \mathbf{z}]$ acting along the $x$-axis \cite{Bra95rpp}. Here, $\mathbf{j}$ is the electric current density and $\mathbf{z}$ is the unit vector in the $z$ direction. If $\mathbf{F}_\mathrm{L}$ exceeds the pinning force $\mathbf{F}_\mathrm{p}$ associated with the local anchoring of vortices to the grooves, the vortex lattice moves in the $x$ direction. Otherwise the vortex lattice remains immobile (pinned) as long as $\mathbf{F}_\mathrm{L} < \mathbf{F}_\mathrm{p}$. These two regimes are stitched together via a nonlinear depinning transition at $\mathbf{F}_\mathrm{L} \thickapprox \mathbf{F}_\mathrm{p}$\,\cite{Bra95rpp}.

A stronger magnon-fluxon interaction is expected for a vortex lattice exhibiting a longer-range order, because in this case vortex rows act as a perfect Bragg grating reflecting spin waves coherently (in phase). To ensure perfect crystallinity of the vortex lattice, the investigated Nb films are of high structural quality, with a very weak intrinsic pinning. For instance, the as-grown Nb films (that is, prior to the milling of grooves) exhibit a flux-flow branch in the current-voltage ($I$-$V$) curves already at very small transport currents\,\cite{Dob19nph}. This is because of the relatively small value of $\mathbf{F}_\mathrm{p}$ associated with the intrinsic pinning in the as-grown Nb films. By contrast, the nanogroove arrays induce a rather strong periodic pinning potential of the washboard type for Abrikosov vortices\,\cite{Dob17pcs}. The stronger pinning in such a landscape allows for the realization of a rather extended range of currents with an immobile (pinned) vortex lattice. Specifically, a transport current applied parallel to the grooves exerts a Lorentz-type force on vortices, making them to overcome the pinning potential barriers associated with the groove slopes. In our studies, one Nb film contains an array of grooves with a symmetric cross-section (sample S) and another film has an asymmetric cross-section (sample A). In sample A, the gentle-groove-slope direction is easy for vortex motion, while a larger transport current is needed to put the vortex lattice into motion in the steep-groove-slope direction.

Magnetic flux quanta are tiny whirls of the supercurrent, producing local magnetic field maxima at the vortex cores \cite{Bra95rpp}. These fields attenuate over a lateral length scale of $2\lambda$, where $\lambda$ is the magnetic penetration depth. With the zero-temperature estimate $\lambda(0) \approx 100\,$nm\,\cite{Gub05prb} and the two-fluid expression\cite{Kim03cry} $\lambda(T) = \lambda(0)[1 - (T/T_c)^4]^{-1/2}$, we obtain $\lambda(8\,\mathrm{K})\approx150$\,nm for our films \cite{Dob12tsf}. From the viewpoint of the spin-wave system, the local magnetic fields created by the vortex lattice constitute a Bragg grating with partial reflection of spin waves at each vortex row. The vortex lattice acts as a magnonic crystal\,\cite{Chu17jpd} featuring bandgaps in the magnon spectrum and leading to characteristic dips in the transmission of spin waves that are directly linked to the wavevector of the fluxon lattice. When the vortex lattice in the S layer is moving, the inelastic scattering of magnons in the F layer on the local magnetic fields emanating from the moving vortex lattice is accompanied by the spin-wave Doppler effect\,\cite{Vla08sci,Kra14pcm,Chu10prb}. The tuning of the vortex lattice velocity by varying the transport current direction and value hence allows for engineering of the bandgaps in the magnon transmission spectrum.
\begin{figure*}[t!]
    \centering
    \includegraphics[width=0.83\linewidth]{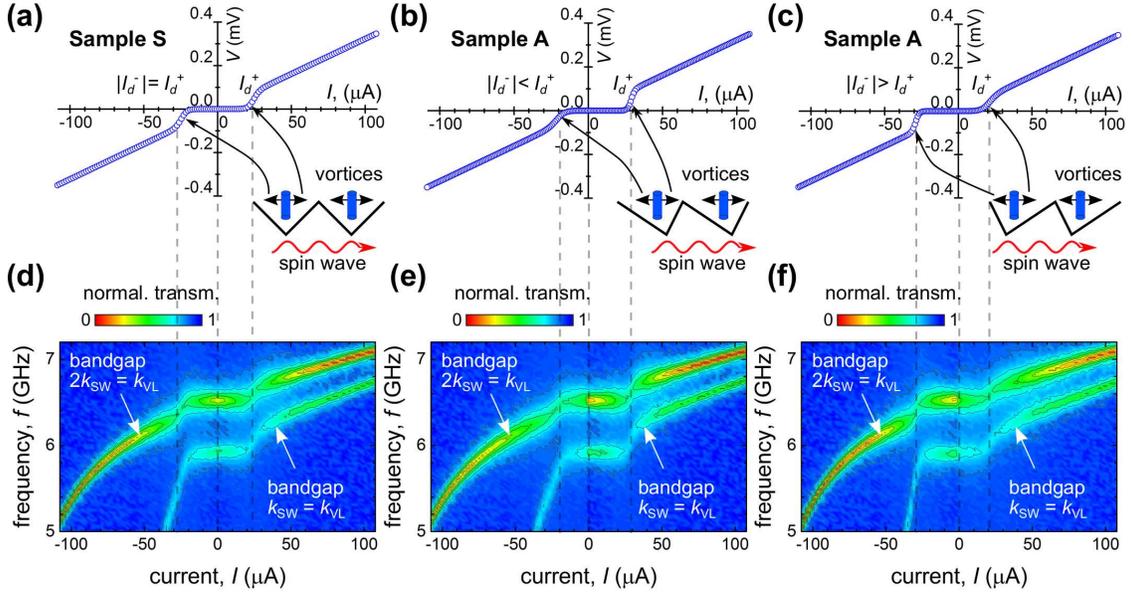}
    \caption{(a) Current-voltage curves for samples S (a) and A (b,c) at $T = 8$\,K and $H_\perp = 7.2$\,mT.
    The mutual orientation of the spin-wave propagation and the motion of vortices in the periodic pinning potential induced by nanogrooves is illustrated in the bottom insets.
    (d-f) Normalized spin-wave transmission as a function of the transport current flowing through the Nb layer.
    The positive current polarity corresponds to co-propagating spin waves and vortex lattice and vice versa.}
    \label{f2}
\end{figure*}

Figure \ref{f2}(a)-(c) presents the $I$-$V$ curves of the Nb microstrips at $T = 8$\,K and $H_\perp = 7.2$\,mT. The $I$-$V$ curve of sample S (Fig. \ref{f2}(a)) is an odd function of $I$, with zero-voltage plateaus at $|I| < |I_\mathrm{d}|\approx22\,\mu$A and a quasi-linear $V(I)$ in the flux-flow regimes at $|I| \gtrsim 30\,\mu$A. Here, $I_\mathrm{d}$ is the depinning current determined by using the $10\,\mu$V voltage criterion. The equality $|I_\mathrm{d}^-| = |I_\mathrm{d}^+|$is associated with the symmetry of the cross-section of the grooves and the equality of the pinning forces $|F_\mathrm{p}^+|=|F_\mathrm{p}^-|$ under current polarity reversal.

The $I$-$V$ curves of sample A for spin waves propagating in the direction of steep and gentle groove slopes are presented in Figs. \ref{f2}(b) and (c), respectively. The vortex motion against the steep groove slope is characterized by a larger depinning current $|I_\mathrm{d}^+|$ as compared to $|I_\mathrm{d}^-|$ for the vortex motion in the gentle-slope direction. The difference in the $I_\mathrm{d}$ values is associated with the difference in the pinning force $F_\mathrm{p} = - \nabla U_\mathrm{p}$ which is larger for the steep-groove-slope direction because of the faster change of the pinning potential $U_\mathrm{p}(x)$ in the direction of vortex motion.

For a static (pinned) vortex lattice at subdepinning currents $|I| < |I_\mathrm{d}|$, spin waves scatter on vortex rows at the Bragg condition $k_\mathrm{SW} = \pi n/a_\mathrm{VL}$, where $k_\mathrm{SW}$ is the spin-wave wavenumber, $n$ is an integer, and $a_\mathrm{VL} =  (2\Phi_0/\sqrt{3}H_\perp)^{1/2}$ is the vortex lattice parameter. The suppression of the microwave transmission at $H_\perp = 7.2$\,mT corresponds to the bandgaps centered at $f_\mathrm{BG1}\approx 6.51$\,GHz and $f_\mathrm{BG2}\approx 5.92$\,GHz. With the general Bragg condition $2a_\mathrm{VL}  = n\lambda_\mathrm{SW}$, which corresponds to $2k_\mathrm{SW}  = nk_\mathrm{VL}$, the frequency positions of the transmission dips correspond\,\cite{Dob19nph} to the wavevectors $k_\mathrm{SW1}  = 5.4\,$rad$/\mu$m and $k_\mathrm{SW1}  = 10.8\,$rad$/\mu$m, i.\,e., correspond to the first- and the second-order Bragg scattering, respectively.

For a moving vortex lattice, the wavevectors of the Bragg-scattered spin waves are modified by $\Delta k_\mathrm{SW}$ because of the translational motion of the vortex lattice. Note, the dispersion of dipolar spin waves in the backward volume magnetostatic spin-wave (BVMSW) geometry (spin waves propagate in the direction (anti)parallel to the direction of the biasing magnetic field $H_{||}$) is characterized by a negative group velocity\,\cite{Chu17jpd}. Accordingly, as the current increases, the bandgap frequencies increase in the case of co-propagating vortex lattice and spin waves and decrease when they are co-propagating, i.\,e., the Doppler effect accompanying the Bragg scattering is reverse\,\cite{Chu10prb}.

We also note that the bandgaps are blurred at the depinning transition (at $I \approx I_\mathrm{d}$). This can be understood as an unavoidable variation in the local pinning forces acting on individual vortices causes their depinning at slightly different current values. Consequently, the vortex lattice loses its long-range order until all vortices have been depinned\,\cite{Bra95rpp}. At larger currents $|I| \gtrsim 30\,\mu$A,  that result in higher vortex velocities, the long-range order in the vortex lattice is recovered. In this free flux-flow regime, the vortex lattice is characterized by a better crystallinity than at the depinning transition\,\cite{Kos94prl}, and the bandgaps are well defined.
\begin{figure*}[t!]
    \centering
    \includegraphics[width=0.85\linewidth]{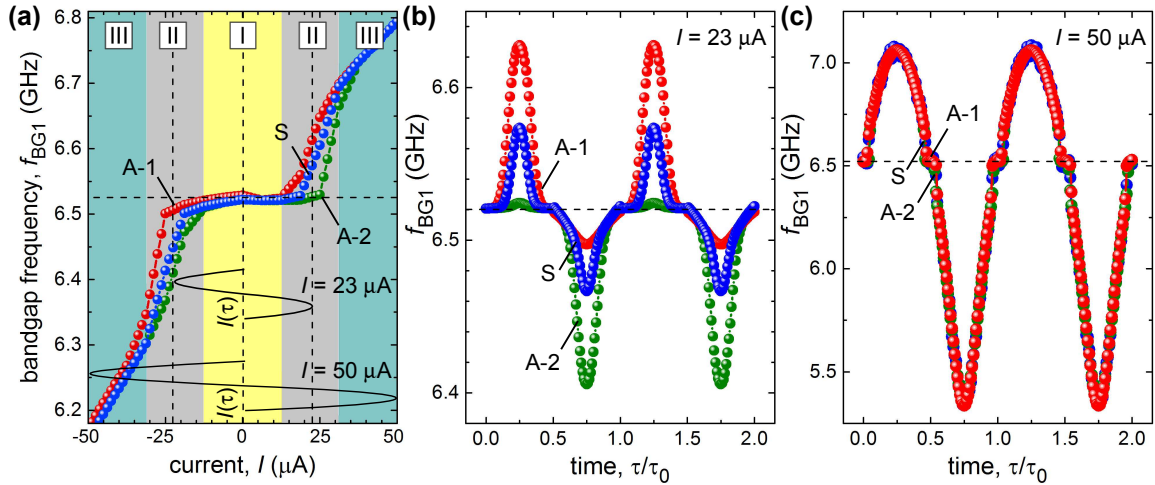}
    \caption{(a) Dependence of the first-order spin-wave bandgap frequency $f_\mathrm{BG1}$ in the Py layer on the transport current applied to the Nb microstrips. The regimes of pinned vortex lattice (I), nonlinear transition (II), and the flux-flow (III) are indicated. In the bottom part of the plot is indicated the amplitude of the ac current resulting in the modulation of $f_\mathrm{BG1}(\tau)$ in panels (b) and (c).}
    \label{f3}
\end{figure*}

The range of currents, at which the bandgap frequencies remain the same as in the static case, is determined by the strength of pinning induced by the nanogrooves. In the flux-flow regime, the nonlinearity of $f_\mathrm{BG1,2}(I)$ is associated with the strong nonlinearity of the dispersion relation $f_\mathrm{BW}(k_\mathrm{SW})$ which has a steeper slope at $k_\mathrm{SW}\rightarrow0$ and goes through a minimum at $k_\mathrm{SW}\approx20$\,rad/$\mu$m, as calculated within the framework of the Kalinikos-Slavin theory adapted for magnonic waveguides\,\cite{Kal86jpc}. Remarkably, a variation of the transport currents from $-100\,\mu$A to $100\,\mu$A allows for tuning the first-order bandgap frequency $f_\mathrm{BG1}$ between about $5.3$ and $7.1$\,GHz. At yet larger currents, the $I$-$V$ curves exhibit a nonlinear upturn just before the transition to the normally conducting state.

The deliberate tuning of the magnon bandgap frequency in the F layer by applying a small-amplitude ac current to the S layer is illustrated in Fig. \ref{f3}. Here, the ac current has an amplitude $I = 23\,\mu$A and a frequency $f_0= 100$\,MHz that corresponds to an ac period $\tau_0=1/f_0=10\,$ns. The time evolution of the first-order bandgap frequency $f_\mathrm{BG1}(\tau)$ for sample S and two orientations of the spin-wave propagation with respect to the steep-groove-slope direction in sample A is presented in Fig. \ref{f3}(b). For all samples, $f_\mathrm{BG1}$ is modulated around $6.51$\,GHz, but the shape of $f_\mathrm{BG1}(\tau)$ decisively depends on the current amplitude and the type of the periodic pinning potential. Specifically, an ac amplitude of $23\,\mu$A is enough to reach the nonlinear transition regime (II) for both ac halfwaves for sample S. Consequently, $f_\mathrm{BG1}$ exhibits almost equally pronounced maxima and minima during positive and negative ac halfwaves. By contrast, the chosen current $23\,\mu$A is smaller than the depinning current $I_\mathrm{d}$ for the steep-slope direction and larger than $I_\mathrm{d}$ for the gentle-slope direction for sample A. Accordingly, $f_\mathrm{BG1}$ is affected strongly during the ac halfway which puts the vortex lattice into motion whereas $f_\mathrm{BG1}$ stays very closely to $6.51$\,GHz in the regime of pinned vortex lattice (I). Finally, in the regime of free flux flow (III), when the pinning induced by the nanogroove arrays no longer plays a crucial role, the curves $f_\mathrm{BG1}(\tau)$ for all samples almost coincide. At the same time, signatures of the different nonlinear regimes remain visible in the $I(\tau)$ dependences as spikes at $f \approx f_\mathrm{BG1}$.

\section{Conclusion}
Summing up, we have demonstrated the scattering of magnons on a lattice of moving Abrikosov fluxons in a superconductor/ferromagnet heterostructure by means of cryogenic microwave spectroscopy. The motion of the vortex lattice results in the Doppler shifts accompanying the Bragg scattering, and the magnitude of the frequency shift defines the position of the bandgaps in the magnon spectrum. Since the bandgap shifts for counter-propagating spin waves are practically anti-symmetric with respect to the current polarity reversal, and the positions of the bandgaps are defined by the applied current for a given strength of the pinning potential, unidirectional spin-wave transport (when the Bragg condition is satisfied for a spin wave propagating in one direction only) can be realized for a spin wave of any frequency. Both symmetric and asymmetric pinning potential landscapes were used in the experiments to realize reciprocal or nonreciprocal motion of the dc-driven fluxon lattice, correspondingly. In the case of asymmetric potential, the vortex motion under current polarity reversal is characterized by different values of the critical depinning current. This difference was used for the realization of a rectified net motion of vortices by an ac current of $23\,\mu$A amplitude and 100\,MHz frequency and for the fast modulation of the bandgap frequencies during an ac current cycle. Finally, the proposed approach can be used for the realization of highly-sensitive dynamically-tunable rf filters with ultra-low energy consumption. Specifically, the application of an ac transport current at a power of 2\,nW in our experiments results in bandgap frequency shifts by up to $\pm100$\,MHz. The tunability of such a filter upon dc current polarity reversal reaches $10$\,GHz/mA. We anticipate a further increase in the frequency of the applied ac current to be possible, yet it requires the understanding of new physical phenomena involving non-elastic multi-magnon scattering processes and their interplay with the vortex-state microwave response in the superconductor \cite{Pom08prb,Dob20pra}.

\section{Materials and methods}
\textbf{Fabrication and properties of the Py magnonic conduits.} The Py waveguides, with a width of $2\,\mu$m, were fabricated by means of a conventional lift-off technique and radiofrequency sputter deposition. To prevent degradation of the magnetic properties on the surface, the Py waveguide was covered with a 5-nm-thick SiO$_2$ capping layer. A reference Py $10\times10\,$mm$^2$ film was characterized by ferromagnetic resonance measurements (FMR) in a broad temperature range ($5$\,K to $300$\,K). From the linewidth of the FMR resonance, a Gilbert damping parameter $\alpha_G$ of about $0.007$ was deduced. The effective magnetization value deduced from fitting the FMR data to the standard Kittel formula is $M_\mathrm{eff}=676$\,kA/m. The Py and Nb layers were deposited on different substrates which were assembled face-to-face for electrical transport measurements.

\textbf{Fabrication and properties of the Nb superconducing strips.} The Nb strips were fabricated by photolithography and Ar etching from epitaxial (110) Nb thin films on a-cut sapphire substrates\,\cite{Dob12tsf}. The Nb films were grown by dc magnetron sputtering in a setup with a base pressure in the $10^{-8}$\,mbar range. During the deposition, the substrate temperature was 850$^\circ$C, the Ar pressure was $4 \times 10^{-3}$\,mbar, and the growth rate was about\,0.5 nm/s. The (110) orientation of the films was inferred from X-ray diffraction measurements. The epitaxy of the films was confirmed by reflection high-energy electron diffraction\,\cite{Dob12tsf}. The as-grown films had a smooth surface with an rms surface roughness of less than $0.2$\,nm, as deduced from atomic force microscopy (AFM) scans in the range $1\,\mu$m$\times1\,\mu$m. The films are in the clean superconducting limit: Their room-temperature-to-10\,K residual resistance ratios are equal to about 30 and the superconducting transition temperatures are $8.92$\,K. The zero-temperature upper critical fields of the Nb films are estimated as $800$\,mT as deduced from fitting the dependence $H_\mathrm{c2}(T)$ to the phenomenological law $H_\mathrm{c2}(T)  =  H_{c2}(0)[1 - (T/T_\mathrm{c})^2]$. After the growth, the Nb films were patterned into the four-probe geometry by photolithography in conjunction with Ar ion etching, forming bridges with a length of $50\,\mu$m and a width of $4\,\mu$m. The Nb microstrips were covered with a 5\,nm-thick SiO$_2$ capping layer and then underwent a nanopatterning step.

\textbf{Fabrication of nanogrooves.} Nanopatterning of the Nb microstrips was done in a high-resolution dual-beam scanning electron microscope (FEI, Nova Nanolab 600) by focused ion beam (FIB) milling\,\cite{Dob18apl}. The nanopatterns are arrays of periodically arranged nanogrooves with symmetric (sample S) and asymmetric (sample A) groove slopes. In the patterning process, the asymmetry of the groove slopes was achieved by defining the grooves in the FIB bitmap file for sample S as a single line for the beam to pass, while a step-wise increasing number of FIB beam passes was assigned to each groove defined as a 5-step ``stair'' for sample A\,\cite{Dob15met}. Due to blurring effects, the symmetric grooves in Sample S have rounded corners while smoothed straight slopes resulted instead of the ``stairs'' in samples A. For all samples the beam parameters were 30\,kV/50\,pA, $1\,\mu$s dwell time and $50$\,nm pitch. The grooves are parallel to the microstrip edges (i.e. to the transport current direction) with a misalignment of less than $0.3^\circ$. Atomic force microscopy (AFM) images of the surfaces of the nanopatterned Nb films are presented in Fig. \ref{f1}(b) and (c). The images were acquired with a Nanosurf easyScan 2 microscope under ambient conditions in non-contact, dynamic force mode. The AFM images reveal a targeted period of $585$\,nm of the nanogroove arrays, with half-depth groove widths of about $50$\,nm for sample S and $150$\,nm for sample A.

\textbf{Electrical transport measurements.} The electrical voltage and microwave transmission measurements were taken in a cryostat equipped with a superconducting solenoid. The microwave signal was provided by a vector network analyzer (VNA) and delivered to/from the heterostructure via semirigid coaxial cables connected to nonmagnetic SMP probes. The forward transmission coefficient (scattering parameter $S_{21}$, associated with the power received at port 2 relative to the power delivered to port 1) was measured by the VNA at the detector port\,\cite{Dob19ami}. Two 100\,nm-thick and 500\,nm-wide Au antennae, with an axis-to-axis distance of $5.5\,\mu$m, were fabricated on the sapphire substrate, spaced by $0.5\,\mu$m on both sides of the edges of the Nb bridge. The microwave signal power was kept at $1\,\mu$W ($-30$\,dBm) being low enough to avoid nonlinear processes. Backward volume magnetostatic spin waves were detected by the second antenna while the excitation frequency was swept in the range from 5 to 8\,GHz. The measurements were taken in an inclined magnetic field $\mathbf{H}_\mathrm{ext}$ with the in-plane component $H_{||}= 59.5$\,mT and the out-of-plane component $H_\perp = 7.2$\,mT kept fixed in all measurements. Herewith, $H_\perp = 7.2$\,mT was chosen as a fundamental matching field for the hexagonal vortex lattice with the $585$\,nm-periodic nanolandscape, see Fig. \ref{f1}(b). In general, any value of $H_{||}$ ensuring magnetization of the Py layer in the in-plane direction could be used, however, we used $H_{||}= 59.5$\,mT to facilitate comparison with previous experiments on plain Nb films\,\cite{Dob19nph}.

\section*{Acknowledgements}
The authors thank Michael Huth for providing access to the laboratory infrastructure, and for valuable discussions. OVD acknowledges the Austrian Science Fund (FWF) for support through Grant No. I 4889 (CurviMag). AVC acknowledges the FWF for support through Grant No. I 4917. Research leading to these results was conducted within the framework of the COST Action CA16218 (NANOCOHYBRI) of the European Cooperation in Science and Technology.


\begin{thebibliography}{53}%
\makeatletter
\providecommand \@ifxundefined [1]{%
 \@ifx{#1\undefined}
}%
\providecommand \@ifnum [1]{%
 \ifnum #1\expandafter \@firstoftwo
 \else \expandafter \@secondoftwo
 \fi
}%
\providecommand \@ifx [1]{%
 \ifx #1\expandafter \@firstoftwo
 \else \expandafter \@secondoftwo
 \fi
}%
\providecommand \natexlab [1]{#1}%
\providecommand \enquote  [1]{``#1''}%
\providecommand \bibnamefont  [1]{#1}%
\providecommand \bibfnamefont [1]{#1}%
\providecommand \citenamefont [1]{#1}%
\providecommand \href@noop [0]{\@secondoftwo}%
\providecommand \href [0]{\begingroup \@sanitize@url \@href}%
\providecommand \@href[1]{\@@startlink{#1}\@@href}%
\providecommand \@@href[1]{\endgroup#1\@@endlink}%
\providecommand \@sanitize@url [0]{\catcode `\\12\catcode `\$12\catcode
  `\&12\catcode `\#12\catcode `\^12\catcode `\_12\catcode `\%12\relax}%
\providecommand \@@startlink[1]{}%
\providecommand \@@endlink[0]{}%
\providecommand \url  [0]{\begingroup\@sanitize@url \@url }%
\providecommand \@url [1]{\endgroup\@href {#1}{\urlprefix }}%
\providecommand \urlprefix  [0]{URL }%
\providecommand \Eprint [0]{\href }%
\providecommand \doibase [0]{http://dx.doi.org/}%
\providecommand \selectlanguage [0]{\@gobble}%
\providecommand \bibinfo  [0]{\@secondoftwo}%
\providecommand \bibfield  [0]{\@secondoftwo}%
\providecommand \translation [1]{[#1]}%
\providecommand \BibitemOpen [0]{}%
\providecommand \bibitemStop [0]{}%
\providecommand \bibitemNoStop [0]{.\EOS\space}%
\providecommand \EOS [0]{\spacefactor3000\relax}%
\providecommand \BibitemShut  [1]{\csname bibitem#1\endcsname}%
\let\auto@bib@innerbib\@empty
\bibitem [{\citenamefont {Barman}\ \emph {et~al.}(2021)\citenamefont {Barman},
  \citenamefont {Gubbiotti}, \citenamefont {Ladak}, \citenamefont {Adeyeye},
  \citenamefont {Krawczyk}, \citenamefont {Gr{\"a}fe}, \citenamefont
  {Adelmann}, \citenamefont {Cotofana}, \citenamefont {Naeemi}, \citenamefont
  {Vasyuchka}, \citenamefont {Hillebrands}, \citenamefont {Nikitov},
  \citenamefont {Yu}, \citenamefont {Grundler}, \citenamefont {Sadovnikov},
  \citenamefont {Grachev}, \citenamefont {Sheshukova}, \citenamefont
  {Duquesne}, \citenamefont {Marangolo}, \citenamefont {Gyorgy}, \citenamefont
  {Porod}, \citenamefont {Demidov}, \citenamefont {Urazhdin}, \citenamefont
  {Demokritov}, \citenamefont {Albisetti}, \citenamefont {Petti}, \citenamefont
  {Bertacco}, \citenamefont {Schulteiss}, \citenamefont {Kruglyak},
  \citenamefont {Poimanov}, \citenamefont {Sahoo}, \citenamefont {Sinha},
  \citenamefont {Yang}, \citenamefont {Muenzenberg}, \citenamefont {Moriyama},
  \citenamefont {Mizukami}, \citenamefont {Landeros}, \citenamefont {Gallardo},
  \citenamefont {Carlotti}, \citenamefont {Kim}, \citenamefont {Stamps},
  \citenamefont {Camley}, \citenamefont {Rana}, \citenamefont {Otani},
  \citenamefont {Yu}, \citenamefont {Yu}, \citenamefont {Bauer}, \citenamefont
  {Back}, \citenamefont {Uhrig}, \citenamefont {Dobrovolskiy}, \citenamefont
  {van Dijken}, \citenamefont {Budinska}, \citenamefont {Qin}, \citenamefont
  {Chumak}, \citenamefont {Khitun}, \citenamefont {Nikonov}, \citenamefont
  {Young}, \citenamefont {Zingsem},\ and\ \citenamefont
  {Winklhofer}}]{Bar21pcm}%
  \BibitemOpen
  \bibfield  {author} {\bibinfo {author} {\bibfnamefont {A.}~\bibnamefont
  {Barman}}, \bibinfo {author} {\bibfnamefont {G.}~\bibnamefont {Gubbiotti}},
  \bibinfo {author} {\bibfnamefont {S.}~\bibnamefont {Ladak}}, \bibinfo
  {author} {\bibfnamefont {A.~O.}\ \bibnamefont {Adeyeye}}, \bibinfo {author}
  {\bibfnamefont {M.}~\bibnamefont {Krawczyk}}, \bibinfo {author}
  {\bibfnamefont {J.}~\bibnamefont {Gr{\"a}fe}}, \bibinfo {author}
  {\bibfnamefont {C.}~\bibnamefont {Adelmann}}, \bibinfo {author}
  {\bibfnamefont {S.}~\bibnamefont {Cotofana}}, \bibinfo {author}
  {\bibfnamefont {A.}~\bibnamefont {Naeemi}}, \bibinfo {author} {\bibfnamefont
  {V.~I.}\ \bibnamefont {Vasyuchka}}, \bibinfo {author} {\bibfnamefont
  {B.}~\bibnamefont {Hillebrands}}, \bibinfo {author} {\bibfnamefont {S.~A.}\
  \bibnamefont {Nikitov}}, \bibinfo {author} {\bibfnamefont {H.}~\bibnamefont
  {Yu}}, \bibinfo {author} {\bibfnamefont {D.}~\bibnamefont {Grundler}},
  \bibinfo {author} {\bibfnamefont {A.}~\bibnamefont {Sadovnikov}}, \bibinfo
  {author} {\bibfnamefont {A.~A.}\ \bibnamefont {Grachev}}, \bibinfo {author}
  {\bibfnamefont {S.~E.}\ \bibnamefont {Sheshukova}}, \bibinfo {author}
  {\bibfnamefont {J.-Y.}\ \bibnamefont {Duquesne}}, \bibinfo {author}
  {\bibfnamefont {M.}~\bibnamefont {Marangolo}}, \bibinfo {author}
  {\bibfnamefont {C.}~\bibnamefont {Gyorgy}}, \bibinfo {author} {\bibfnamefont
  {W.}~\bibnamefont {Porod}}, \bibinfo {author} {\bibfnamefont {V.~E.}\
  \bibnamefont {Demidov}}, \bibinfo {author} {\bibfnamefont {S.}~\bibnamefont
  {Urazhdin}}, \bibinfo {author} {\bibfnamefont {S.}~\bibnamefont
  {Demokritov}}, \bibinfo {author} {\bibfnamefont {E.}~\bibnamefont
  {Albisetti}}, \bibinfo {author} {\bibfnamefont {D.}~\bibnamefont {Petti}},
  \bibinfo {author} {\bibfnamefont {R.}~\bibnamefont {Bertacco}}, \bibinfo
  {author} {\bibfnamefont {H.}~\bibnamefont {Schulteiss}}, \bibinfo {author}
  {\bibfnamefont {V.~V.}\ \bibnamefont {Kruglyak}}, \bibinfo {author}
  {\bibfnamefont {V.~D.}\ \bibnamefont {Poimanov}}, \bibinfo {author}
  {\bibfnamefont {A.~K.}\ \bibnamefont {Sahoo}}, \bibinfo {author}
  {\bibfnamefont {J.}~\bibnamefont {Sinha}}, \bibinfo {author} {\bibfnamefont
  {H.}~\bibnamefont {Yang}}, \bibinfo {author} {\bibfnamefont {M.}~\bibnamefont
  {Muenzenberg}}, \bibinfo {author} {\bibfnamefont {T.}~\bibnamefont
  {Moriyama}}, \bibinfo {author} {\bibfnamefont {S.}~\bibnamefont {Mizukami}},
  \bibinfo {author} {\bibfnamefont {P.}~\bibnamefont {Landeros}}, \bibinfo
  {author} {\bibfnamefont {R.~A.}\ \bibnamefont {Gallardo}}, \bibinfo {author}
  {\bibfnamefont {G.}~\bibnamefont {Carlotti}}, \bibinfo {author}
  {\bibfnamefont {J.-V.}\ \bibnamefont {Kim}}, \bibinfo {author} {\bibfnamefont
  {R.~L.}\ \bibnamefont {Stamps}}, \bibinfo {author} {\bibfnamefont {R.~E.}\
  \bibnamefont {Camley}}, \bibinfo {author} {\bibfnamefont {B.}~\bibnamefont
  {Rana}}, \bibinfo {author} {\bibfnamefont {Y.}~\bibnamefont {Otani}},
  \bibinfo {author} {\bibfnamefont {W.}~\bibnamefont {Yu}}, \bibinfo {author}
  {\bibfnamefont {T.}~\bibnamefont {Yu}}, \bibinfo {author} {\bibfnamefont
  {G.~E.~W.}\ \bibnamefont {Bauer}}, \bibinfo {author} {\bibfnamefont {C.~H.}\
  \bibnamefont {Back}}, \bibinfo {author} {\bibfnamefont {G.~S.}\ \bibnamefont
  {Uhrig}}, \bibinfo {author} {\bibfnamefont {O.~V.}\ \bibnamefont
  {Dobrovolskiy}}, \bibinfo {author} {\bibfnamefont {S.}~\bibnamefont {van
  Dijken}}, \bibinfo {author} {\bibfnamefont {B.}~\bibnamefont {Budinska}},
  \bibinfo {author} {\bibfnamefont {H.}~\bibnamefont {Qin}}, \bibinfo {author}
  {\bibfnamefont {A.}~\bibnamefont {Chumak}}, \bibinfo {author} {\bibfnamefont
  {A.}~\bibnamefont {Khitun}}, \bibinfo {author} {\bibfnamefont {D.~E.}\
  \bibnamefont {Nikonov}}, \bibinfo {author} {\bibfnamefont {I.~A.}\
  \bibnamefont {Young}}, \bibinfo {author} {\bibfnamefont {B.}~\bibnamefont
  {Zingsem}}, \ and\ \bibinfo {author} {\bibfnamefont {M.}~\bibnamefont
  {Winklhofer}},\ }\href {\doibase https://doi.org/10.1088/1361-648X/abec1a}
  {\bibfield  {journal} {\bibinfo  {journal} {J. Phys.: Condens. Matt.}\ }
  (\bibinfo {year} {2021}),\
  https://doi.org/10.1088/1361-648X/abec1a}\BibitemShut {NoStop}%
\bibitem [{\citenamefont {Mahmoud}\ \emph {et~al.}(2020)\citenamefont
  {Mahmoud}, \citenamefont {Ciubotaru}, \citenamefont {Vanderveken},
  \citenamefont {Chumak}, \citenamefont {Hamdioui}, \citenamefont {Adelmann},\
  and\ \citenamefont {Cotofana}}]{Mah20jap}%
  \BibitemOpen
  \bibfield  {author} {\bibinfo {author} {\bibfnamefont {A.}~\bibnamefont
  {Mahmoud}}, \bibinfo {author} {\bibfnamefont {F.}~\bibnamefont {Ciubotaru}},
  \bibinfo {author} {\bibfnamefont {F.}~\bibnamefont {Vanderveken}}, \bibinfo
  {author} {\bibfnamefont {A.~V.}\ \bibnamefont {Chumak}}, \bibinfo {author}
  {\bibfnamefont {S.}~\bibnamefont {Hamdioui}}, \bibinfo {author}
  {\bibfnamefont {C.}~\bibnamefont {Adelmann}}, \ and\ \bibinfo {author}
  {\bibfnamefont {S.}~\bibnamefont {Cotofana}},\ }\href {\doibase
  10.1063/5.0019328} {\bibfield  {journal} {\bibinfo  {journal} {J. Appl.
  Phys.}\ }\textbf {\bibinfo {volume} {128}},\ \bibinfo {pages} {161101}
  (\bibinfo {year} {2020})}\BibitemShut {NoStop}%
\bibitem [{\citenamefont {Wang}\ \emph {et~al.}(2020)\citenamefont {Wang},
  \citenamefont {Kewenig}, \citenamefont {Schneider}, \citenamefont {Verba},
  \citenamefont {Kohl}, \citenamefont {Heinz}, \citenamefont {Geilen},
  \citenamefont {Mohseni}, \citenamefont {L{\"a}gel}, \citenamefont
  {Ciubotaru}, \citenamefont {Adelmann}, \citenamefont {Dubs}, \citenamefont
  {Cotofana}, \citenamefont {Dobrovolskiy}, \citenamefont {Br{\"a}cher},
  \citenamefont {Pirro},\ and\ \citenamefont {Chumak}}]{Wan20nel}%
  \BibitemOpen
  \bibfield  {author} {\bibinfo {author} {\bibfnamefont {Q.}~\bibnamefont
  {Wang}}, \bibinfo {author} {\bibfnamefont {M.}~\bibnamefont {Kewenig}},
  \bibinfo {author} {\bibfnamefont {M.}~\bibnamefont {Schneider}}, \bibinfo
  {author} {\bibfnamefont {R.}~\bibnamefont {Verba}}, \bibinfo {author}
  {\bibfnamefont {F.}~\bibnamefont {Kohl}}, \bibinfo {author} {\bibfnamefont
  {B.}~\bibnamefont {Heinz}}, \bibinfo {author} {\bibfnamefont
  {M.}~\bibnamefont {Geilen}}, \bibinfo {author} {\bibfnamefont
  {M.}~\bibnamefont {Mohseni}}, \bibinfo {author} {\bibfnamefont
  {B.}~\bibnamefont {L{\"a}gel}}, \bibinfo {author} {\bibfnamefont
  {F.}~\bibnamefont {Ciubotaru}}, \bibinfo {author} {\bibfnamefont
  {C.}~\bibnamefont {Adelmann}}, \bibinfo {author} {\bibfnamefont
  {C.}~\bibnamefont {Dubs}}, \bibinfo {author} {\bibfnamefont {S.~D.}\
  \bibnamefont {Cotofana}}, \bibinfo {author} {\bibfnamefont {O.~V.}\
  \bibnamefont {Dobrovolskiy}}, \bibinfo {author} {\bibfnamefont
  {T.}~\bibnamefont {Br{\"a}cher}}, \bibinfo {author} {\bibfnamefont
  {P.}~\bibnamefont {Pirro}}, \ and\ \bibinfo {author} {\bibfnamefont {A.~V.}\
  \bibnamefont {Chumak}},\ }\href {\doibase 10.1038/s41928-020-00485-6}
  {\bibfield  {journal} {\bibinfo  {journal} {Nature Electron.}\ }\textbf
  {\bibinfo {volume} {3}},\ \bibinfo {pages} {765} (\bibinfo {year}
  {2020})}\BibitemShut {NoStop}%
\bibitem [{\citenamefont {Dobrovolskiy}\ \emph
  {et~al.}(2019{\natexlab{a}})\citenamefont {Dobrovolskiy}, \citenamefont
  {Sachser}, \citenamefont {Bunyaev}, \citenamefont {Navas}, \citenamefont
  {Bevz}, \citenamefont {Zelent}, \citenamefont {Smigaj}, \citenamefont
  {Rychly}, \citenamefont {Krawczyk}, \citenamefont {Vovk}, \citenamefont
  {Huth},\ and\ \citenamefont {Kakazei}}]{Dob19ami}%
  \BibitemOpen
  \bibfield  {author} {\bibinfo {author} {\bibfnamefont {O.~V.}\ \bibnamefont
  {Dobrovolskiy}}, \bibinfo {author} {\bibfnamefont {R.}~\bibnamefont
  {Sachser}}, \bibinfo {author} {\bibfnamefont {S.~A.}\ \bibnamefont
  {Bunyaev}}, \bibinfo {author} {\bibfnamefont {D.}~\bibnamefont {Navas}},
  \bibinfo {author} {\bibfnamefont {V.~M.}\ \bibnamefont {Bevz}}, \bibinfo
  {author} {\bibfnamefont {M.}~\bibnamefont {Zelent}}, \bibinfo {author}
  {\bibfnamefont {W.}~\bibnamefont {Smigaj}}, \bibinfo {author} {\bibfnamefont
  {J.}~\bibnamefont {Rychly}}, \bibinfo {author} {\bibfnamefont
  {M.}~\bibnamefont {Krawczyk}}, \bibinfo {author} {\bibfnamefont {R.~V.}\
  \bibnamefont {Vovk}}, \bibinfo {author} {\bibfnamefont {M.}~\bibnamefont
  {Huth}}, \ and\ \bibinfo {author} {\bibfnamefont {G.~N.}\ \bibnamefont
  {Kakazei}},\ }\href {\doibase 10.1021/acsami.9b02717} {\bibfield  {journal}
  {\bibinfo  {journal} {ACS Appl. Mater. Interf.}\ }\textbf {\bibinfo {volume}
  {11}},\ \bibinfo {pages} {17654} (\bibinfo {year}
  {2019}{\natexlab{a}})}\BibitemShut {NoStop}%
\bibitem [{\citenamefont {Papp}\ \emph {et~al.}()\citenamefont {Papp},
  \citenamefont {Porod},\ and\ \citenamefont {Csaba}}]{Pap20arx}%
  \BibitemOpen
  \bibfield  {author} {\bibinfo {author} {\bibfnamefont {A.}~\bibnamefont
  {Papp}}, \bibinfo {author} {\bibfnamefont {W.}~\bibnamefont {Porod}}, \ and\
  \bibinfo {author} {\bibfnamefont {G.}~\bibnamefont {Csaba}},\ }\href@noop {}
  {\bibinfo  {journal} {ArXiv:2012.04594}\ }\BibitemShut {NoStop}%
\bibitem [{\citenamefont {Wang}\ \emph {et~al.}(2021)\citenamefont {Wang},
  \citenamefont {Chumak},\ and\ \citenamefont {Pirro}}]{Wan21nac}%
  \BibitemOpen
\bibfield  {journal} {  }\bibfield  {author} {\bibinfo {author} {\bibfnamefont
  {Q.}~\bibnamefont {Wang}}, \bibinfo {author} {\bibfnamefont {A.~V.}\
  \bibnamefont {Chumak}}, \ and\ \bibinfo {author} {\bibfnamefont
  {P.}~\bibnamefont {Pirro}},\ }\href {\doibase 10.1038/s41467-021-22897-4}
  {\bibfield  {journal} {\bibinfo  {journal} {Nat. Commun.}\ }\textbf {\bibinfo
  {volume} {12}},\ \bibinfo {pages} {2636} (\bibinfo {year}
  {2021})}\BibitemShut {NoStop}%
\bibitem [{\citenamefont {Dobrovolskiy}\ \emph
  {et~al.}(2021{\natexlab{a}})\citenamefont {Dobrovolskiy}, \citenamefont
  {Vovk}, \citenamefont {Bondarenko}, \citenamefont {Bunyaev}, \citenamefont
  {Lamb-Camarena}, \citenamefont {Zenbaa}, \citenamefont {Sachser},
  \citenamefont {Barth}, \citenamefont {Guslienko}, \citenamefont {Chumak},
  \citenamefont {Huth},\ and\ \citenamefont {Kakazei}}]{Dob21apl}%
  \BibitemOpen
  \bibfield  {author} {\bibinfo {author} {\bibfnamefont {O.~V.}\ \bibnamefont
  {Dobrovolskiy}}, \bibinfo {author} {\bibfnamefont {N.~R.}\ \bibnamefont
  {Vovk}}, \bibinfo {author} {\bibfnamefont {A.~V.}\ \bibnamefont
  {Bondarenko}}, \bibinfo {author} {\bibfnamefont {S.~A.}\ \bibnamefont
  {Bunyaev}}, \bibinfo {author} {\bibfnamefont {S.}~\bibnamefont
  {Lamb-Camarena}}, \bibinfo {author} {\bibfnamefont {N.}~\bibnamefont
  {Zenbaa}}, \bibinfo {author} {\bibfnamefont {R.}~\bibnamefont {Sachser}},
  \bibinfo {author} {\bibfnamefont {S.}~\bibnamefont {Barth}}, \bibinfo
  {author} {\bibfnamefont {K.~Y.}\ \bibnamefont {Guslienko}}, \bibinfo {author}
  {\bibfnamefont {A.~V.}\ \bibnamefont {Chumak}}, \bibinfo {author}
  {\bibfnamefont {M.}~\bibnamefont {Huth}}, \ and\ \bibinfo {author}
  {\bibfnamefont {G.~N.}\ \bibnamefont {Kakazei}},\ }\href {\doibase
  10.1063/5.0044325} {\bibfield  {journal} {\bibinfo  {journal} {Appl. Phys.
  Lett.}\ }\textbf {\bibinfo {volume} {118}},\ \bibinfo {pages} {132405}
  (\bibinfo {year} {2021}{\natexlab{a}})}\BibitemShut {NoStop}%
\bibitem [{\citenamefont {Heinz}\ \emph {et~al.}(2020)\citenamefont {Heinz},
  \citenamefont {Br{\"a}cher}, \citenamefont {Schneider}, \citenamefont {Wang},
  \citenamefont {L{\"a}gel}, \citenamefont {Friedel}, \citenamefont
  {Breitbach}, \citenamefont {Steinert}, \citenamefont {Meyer}, \citenamefont
  {Kewenig}, \citenamefont {Dubs}, \citenamefont {Pirro},\ and\ \citenamefont
  {Chumak}}]{Hei20nal}%
  \BibitemOpen
  \bibfield  {author} {\bibinfo {author} {\bibfnamefont {B.}~\bibnamefont
  {Heinz}}, \bibinfo {author} {\bibfnamefont {T.}~\bibnamefont {Br{\"a}cher}},
  \bibinfo {author} {\bibfnamefont {M.}~\bibnamefont {Schneider}}, \bibinfo
  {author} {\bibfnamefont {Q.}~\bibnamefont {Wang}}, \bibinfo {author}
  {\bibfnamefont {B.}~\bibnamefont {L{\"a}gel}}, \bibinfo {author}
  {\bibfnamefont {A.~M.}\ \bibnamefont {Friedel}}, \bibinfo {author}
  {\bibfnamefont {D.}~\bibnamefont {Breitbach}}, \bibinfo {author}
  {\bibfnamefont {S.}~\bibnamefont {Steinert}}, \bibinfo {author}
  {\bibfnamefont {T.}~\bibnamefont {Meyer}}, \bibinfo {author} {\bibfnamefont
  {M.}~\bibnamefont {Kewenig}}, \bibinfo {author} {\bibfnamefont
  {C.}~\bibnamefont {Dubs}}, \bibinfo {author} {\bibfnamefont {P.}~\bibnamefont
  {Pirro}}, \ and\ \bibinfo {author} {\bibfnamefont {A.~V.}\ \bibnamefont
  {Chumak}},\ }\href {\doibase 10.1021/acs.nanolett.0c00657} {\bibfield
  {journal} {\bibinfo  {journal} {Nano Lett.}\ }\textbf {\bibinfo {volume}
  {20}},\ \bibinfo {pages} {4220} (\bibinfo {year} {2020})}\BibitemShut
  {NoStop}%
\bibitem [{\citenamefont {Che}\ \emph {et~al.}(2020)\citenamefont {Che},
  \citenamefont {Baumgaertl}, \citenamefont {K{\'u}kol'ov{\'a}}, \citenamefont
  {Dubs},\ and\ \citenamefont {Grundler}}]{Che20nac}%
  \BibitemOpen
  \bibfield  {author} {\bibinfo {author} {\bibfnamefont {P.}~\bibnamefont
  {Che}}, \bibinfo {author} {\bibfnamefont {K.}~\bibnamefont {Baumgaertl}},
  \bibinfo {author} {\bibfnamefont {A.}~\bibnamefont {K{\'u}kol'ov{\'a}}},
  \bibinfo {author} {\bibfnamefont {C.}~\bibnamefont {Dubs}}, \ and\ \bibinfo
  {author} {\bibfnamefont {D.}~\bibnamefont {Grundler}},\ }\href {\doibase
  10.1038/s41467-020-15265-1} {\bibfield  {journal} {\bibinfo  {journal} {Nat.
  Commun.}\ }\textbf {\bibinfo {volume} {11}},\ \bibinfo {pages} {1445}
  (\bibinfo {year} {2020})}\BibitemShut {NoStop}%
\bibitem [{\citenamefont {Baumgaertl}\ and\ \citenamefont
  {Grundler}(2021)}]{Bau21apl}%
  \BibitemOpen
  \bibfield  {author} {\bibinfo {author} {\bibfnamefont {K.}~\bibnamefont
  {Baumgaertl}}\ and\ \bibinfo {author} {\bibfnamefont {D.}~\bibnamefont
  {Grundler}},\ }\href {\doibase 10.1063/5.0048825} {\bibfield  {journal}
  {\bibinfo  {journal} {Appl. Phys. Lett.}\ }\textbf {\bibinfo {volume}
  {118}},\ \bibinfo {pages} {162402} (\bibinfo {year} {2021})}\BibitemShut
  {NoStop}%
\bibitem [{\citenamefont {Liu}\ \emph {et~al.}(2018)\citenamefont {Liu},
  \citenamefont {Chen}, \citenamefont {Liu}, \citenamefont {Heimbach},
  \citenamefont {Yu}, \citenamefont {Xiao}, \citenamefont {Hu}, \citenamefont
  {Liu}, \citenamefont {Chang}, \citenamefont {Stueckler}, \citenamefont
  {Zhang}, \citenamefont {Zhang}, \citenamefont {Gao}, \citenamefont {Liao},
  \citenamefont {Yu}, \citenamefont {Xia}, \citenamefont {Lei}, \citenamefont
  {Zhao},\ and\ \citenamefont {Wu}}]{Liu18nac}%
  \BibitemOpen
  \bibfield  {author} {\bibinfo {author} {\bibfnamefont {C.}~\bibnamefont
  {Liu}}, \bibinfo {author} {\bibfnamefont {J.}~\bibnamefont {Chen}}, \bibinfo
  {author} {\bibfnamefont {T.}~\bibnamefont {Liu}}, \bibinfo {author}
  {\bibfnamefont {F.}~\bibnamefont {Heimbach}}, \bibinfo {author}
  {\bibfnamefont {H.}~\bibnamefont {Yu}}, \bibinfo {author} {\bibfnamefont
  {Y.}~\bibnamefont {Xiao}}, \bibinfo {author} {\bibfnamefont {J.}~\bibnamefont
  {Hu}}, \bibinfo {author} {\bibfnamefont {M.}~\bibnamefont {Liu}}, \bibinfo
  {author} {\bibfnamefont {H.}~\bibnamefont {Chang}}, \bibinfo {author}
  {\bibfnamefont {S.}~\bibnamefont {Stueckler}, \bibfnamefont {T.and~Tu}},
  \bibinfo {author} {\bibfnamefont {Y.}~\bibnamefont {Zhang}}, \bibinfo
  {author} {\bibfnamefont {Y.}~\bibnamefont {Zhang}}, \bibinfo {author}
  {\bibfnamefont {P.}~\bibnamefont {Gao}}, \bibinfo {author} {\bibfnamefont
  {Z.}~\bibnamefont {Liao}}, \bibinfo {author} {\bibfnamefont {D.}~\bibnamefont
  {Yu}}, \bibinfo {author} {\bibfnamefont {K.}~\bibnamefont {Xia}}, \bibinfo
  {author} {\bibfnamefont {N.}~\bibnamefont {Lei}}, \bibinfo {author}
  {\bibfnamefont {W.}~\bibnamefont {Zhao}}, \ and\ \bibinfo {author}
  {\bibfnamefont {M.}~\bibnamefont {Wu}},\ }\href {\doibase
  10.1038/s41467-018-03199-8} {\bibfield  {journal} {\bibinfo  {journal} {Nat.
  Commun.}\ }\textbf {\bibinfo {volume} {9}},\ \bibinfo {pages} {738} (\bibinfo
  {year} {2018})}\BibitemShut {NoStop}%
\bibitem [{\citenamefont {Dobrovolskiy}\ \emph
  {et~al.}(2021{\natexlab{b}})\citenamefont {Dobrovolskiy}, \citenamefont
  {Wang}, \citenamefont {Vodolazov}, \citenamefont {Budinska}, \citenamefont
  {Sachser}, \citenamefont {Chumak}, \citenamefont {Huth},\ and\ \citenamefont
  {Buzdin}}]{Dob21arx}%
  \BibitemOpen
  \bibfield  {author} {\bibinfo {author} {\bibfnamefont {O.~V.}\ \bibnamefont
  {Dobrovolskiy}}, \bibinfo {author} {\bibfnamefont {Q.}~\bibnamefont {Wang}},
  \bibinfo {author} {\bibfnamefont {D.~Y.}\ \bibnamefont {Vodolazov}}, \bibinfo
  {author} {\bibfnamefont {B.}~\bibnamefont {Budinska}}, \bibinfo {author}
  {\bibfnamefont {R.}~\bibnamefont {Sachser}}, \bibinfo {author} {\bibfnamefont
  {A.~V.}\ \bibnamefont {Chumak}}, \bibinfo {author} {\bibfnamefont
  {M.}~\bibnamefont {Huth}}, \ and\ \bibinfo {author} {\bibfnamefont {A.~I.}\
  \bibnamefont {Buzdin}},\ }\href@noop {} {\bibfield  {journal} {\bibinfo
  {journal} {arXiv:2103.10156}\ } (\bibinfo {year}
  {2021}{\natexlab{b}})}\BibitemShut {NoStop}%
\bibitem [{\citenamefont {Demokritov}\ \emph {et~al.}(2006)\citenamefont
  {Demokritov}, \citenamefont {Demidov}, \citenamefont {Dzyapko}, \citenamefont
  {Melkov}, \citenamefont {Serga}, \citenamefont {Hillebrands},\ and\
  \citenamefont {Slavin}}]{Dem06nat}%
  \BibitemOpen
  \bibfield  {author} {\bibinfo {author} {\bibfnamefont {S.~O.}\ \bibnamefont
  {Demokritov}}, \bibinfo {author} {\bibfnamefont {V.~E.}\ \bibnamefont
  {Demidov}}, \bibinfo {author} {\bibfnamefont {O.}~\bibnamefont {Dzyapko}},
  \bibinfo {author} {\bibfnamefont {G.~A.}\ \bibnamefont {Melkov}}, \bibinfo
  {author} {\bibfnamefont {A.~A.}\ \bibnamefont {Serga}}, \bibinfo {author}
  {\bibfnamefont {B.}~\bibnamefont {Hillebrands}}, \ and\ \bibinfo {author}
  {\bibfnamefont {A.~N.}\ \bibnamefont {Slavin}},\ }\href
  {http://dx.doi.org/10.1038/nature05117} {\bibfield  {journal} {\bibinfo
  {journal} {Nature}\ }\textbf {\bibinfo {volume} {443}},\ \bibinfo {pages}
  {430} (\bibinfo {year} {2006})}\BibitemShut {NoStop}%
\bibitem [{\citenamefont {Schneider}\ \emph {et~al.}(2020)\citenamefont
  {Schneider}, \citenamefont {Br{\"a}cher}, \citenamefont {Breitbach},
  \citenamefont {Lauer}, \citenamefont {Pirro}, \citenamefont {Bozhko},
  \citenamefont {Musiienko-Shmarova}, \citenamefont {Heinz}, \citenamefont
  {Wang}, \citenamefont {Meyer}, \citenamefont {Heussner}, \citenamefont
  {Keller}, \citenamefont {Papaioannou}, \citenamefont {L{\"a}gel},
  \citenamefont {L{\"o}ber}, \citenamefont {Dubs}, \citenamefont {Slavin},
  \citenamefont {Tiberkevich}, \citenamefont {Serga}, \citenamefont
  {Hillebrands},\ and\ \citenamefont {Chumak}}]{Sch20nan}%
  \BibitemOpen
  \bibfield  {author} {\bibinfo {author} {\bibfnamefont {M.}~\bibnamefont
  {Schneider}}, \bibinfo {author} {\bibfnamefont {T.}~\bibnamefont
  {Br{\"a}cher}}, \bibinfo {author} {\bibfnamefont {D.}~\bibnamefont
  {Breitbach}}, \bibinfo {author} {\bibfnamefont {V.}~\bibnamefont {Lauer}},
  \bibinfo {author} {\bibfnamefont {P.}~\bibnamefont {Pirro}}, \bibinfo
  {author} {\bibfnamefont {D.~A.}\ \bibnamefont {Bozhko}}, \bibinfo {author}
  {\bibfnamefont {H.~Y.}\ \bibnamefont {Musiienko-Shmarova}}, \bibinfo {author}
  {\bibfnamefont {B.}~\bibnamefont {Heinz}}, \bibinfo {author} {\bibfnamefont
  {Q.}~\bibnamefont {Wang}}, \bibinfo {author} {\bibfnamefont {T.}~\bibnamefont
  {Meyer}}, \bibinfo {author} {\bibfnamefont {F.}~\bibnamefont {Heussner}},
  \bibinfo {author} {\bibfnamefont {S.}~\bibnamefont {Keller}}, \bibinfo
  {author} {\bibfnamefont {E.~T.}\ \bibnamefont {Papaioannou}}, \bibinfo
  {author} {\bibfnamefont {B.}~\bibnamefont {L{\"a}gel}}, \bibinfo {author}
  {\bibfnamefont {T.}~\bibnamefont {L{\"o}ber}}, \bibinfo {author}
  {\bibfnamefont {C.}~\bibnamefont {Dubs}}, \bibinfo {author} {\bibfnamefont
  {A.~N.}\ \bibnamefont {Slavin}}, \bibinfo {author} {\bibfnamefont {V.~S.}\
  \bibnamefont {Tiberkevich}}, \bibinfo {author} {\bibfnamefont {A.~A.}\
  \bibnamefont {Serga}}, \bibinfo {author} {\bibfnamefont {B.}~\bibnamefont
  {Hillebrands}}, \ and\ \bibinfo {author} {\bibfnamefont {A.~V.}\ \bibnamefont
  {Chumak}},\ }\href {\doibase 10.1038/s41565-020-0671-z} {\bibfield  {journal}
  {\bibinfo  {journal} {Nat. Nanotechnol.}\ }\textbf {\bibinfo {volume} {15}},\
  \bibinfo {pages} {457} (\bibinfo {year} {2020})}\BibitemShut {NoStop}%
\bibitem [{\citenamefont {Schneider}\ \emph {et~al.}(2021)\citenamefont
  {Schneider}, \citenamefont {Breitbach}, \citenamefont {Serha}, \citenamefont
  {Wang}, \citenamefont {Serga}, \citenamefont {Slavin}, \citenamefont
  {Tiberkevich}, \citenamefont {Heinz}, \citenamefont {L\"agel}, \citenamefont
  {Br\"acher}, \citenamefont {Dubs}, \citenamefont {Knauer}, \citenamefont
  {Dobrovolskiy}, \citenamefont {Pirro}, \citenamefont {Hillebrands},\ and\
  \citenamefont {Chumak}}]{Sch21arx}%
  \BibitemOpen
  \bibfield  {author} {\bibinfo {author} {\bibfnamefont {M.}~\bibnamefont
  {Schneider}}, \bibinfo {author} {\bibfnamefont {D.}~\bibnamefont
  {Breitbach}}, \bibinfo {author} {\bibfnamefont {R.}~\bibnamefont {Serha}},
  \bibinfo {author} {\bibfnamefont {Q.}~\bibnamefont {Wang}}, \bibinfo {author}
  {\bibfnamefont {A.~A.}\ \bibnamefont {Serga}}, \bibinfo {author}
  {\bibfnamefont {A.~N.}\ \bibnamefont {Slavin}}, \bibinfo {author}
  {\bibfnamefont {V.~S.}\ \bibnamefont {Tiberkevich}}, \bibinfo {author}
  {\bibfnamefont {B.}~\bibnamefont {Heinz}}, \bibinfo {author} {\bibfnamefont
  {B.}~\bibnamefont {L\"agel}}, \bibinfo {author} {\bibfnamefont
  {T.}~\bibnamefont {Br\"acher}}, \bibinfo {author} {\bibfnamefont
  {C.}~\bibnamefont {Dubs}}, \bibinfo {author} {\bibfnamefont {S.}~\bibnamefont
  {Knauer}}, \bibinfo {author} {\bibfnamefont {O.~V.}\ \bibnamefont
  {Dobrovolskiy}}, \bibinfo {author} {\bibfnamefont {P.}~\bibnamefont {Pirro}},
  \bibinfo {author} {\bibfnamefont {B.}~\bibnamefont {Hillebrands}}, \ and\
  \bibinfo {author} {\bibfnamefont {A.~V.}\ \bibnamefont {Chumak}},\
  }\href@noop {} {\bibfield  {journal} {\bibinfo  {journal} {arXiv:2102.13481}\
  } (\bibinfo {year} {2021})}\BibitemShut {NoStop}%
\bibitem [{\citenamefont {Lachance-Quirion}\ \emph {et~al.}(2020)\citenamefont
  {Lachance-Quirion}, \citenamefont {Wolski}, \citenamefont {Tabuchi},
  \citenamefont {Kono}, \citenamefont {Usami},\ and\ \citenamefont
  {Nakamura}}]{Lac20sci}%
  \BibitemOpen
  \bibfield  {author} {\bibinfo {author} {\bibfnamefont {D.}~\bibnamefont
  {Lachance-Quirion}}, \bibinfo {author} {\bibfnamefont {S.~P.}\ \bibnamefont
  {Wolski}}, \bibinfo {author} {\bibfnamefont {Y.}~\bibnamefont {Tabuchi}},
  \bibinfo {author} {\bibfnamefont {S.}~\bibnamefont {Kono}}, \bibinfo {author}
  {\bibfnamefont {K.}~\bibnamefont {Usami}}, \ and\ \bibinfo {author}
  {\bibfnamefont {Y.}~\bibnamefont {Nakamura}},\ }\href {\doibase
  10.1126/science.aaz9236} {\bibfield  {journal} {\bibinfo  {journal}
  {Science}\ }\textbf {\bibinfo {volume} {367}},\ \bibinfo {pages} {425}
  (\bibinfo {year} {2020})}\BibitemShut {NoStop}%
\bibitem [{\citenamefont {Rana}\ \emph {et~al.}(2019)\citenamefont {Rana},
  \citenamefont {Choudhury}, \citenamefont {Miura}, \citenamefont {Takahashi},
  \citenamefont {Barman},\ and\ \citenamefont {Otani}}]{Ran19prb}%
  \BibitemOpen
  \bibfield  {author} {\bibinfo {author} {\bibfnamefont {B.}~\bibnamefont
  {Rana}}, \bibinfo {author} {\bibfnamefont {S.}~\bibnamefont {Choudhury}},
  \bibinfo {author} {\bibfnamefont {K.}~\bibnamefont {Miura}}, \bibinfo
  {author} {\bibfnamefont {H.}~\bibnamefont {Takahashi}}, \bibinfo {author}
  {\bibfnamefont {A.}~\bibnamefont {Barman}}, \ and\ \bibinfo {author}
  {\bibfnamefont {Y.}~\bibnamefont {Otani}},\ }\href {\doibase
  10.1103/PhysRevB.100.224412} {\bibfield  {journal} {\bibinfo  {journal}
  {Phys. Rev. B}\ }\textbf {\bibinfo {volume} {100}},\ \bibinfo {pages}
  {224412} (\bibinfo {year} {2019})}\BibitemShut {NoStop}%
\bibitem [{\citenamefont {Chen}\ \emph {et~al.}(2017)\citenamefont {Chen},
  \citenamefont {Lee}, \citenamefont {Verba}, \citenamefont {Katine},
  \citenamefont {Barsukov}, \citenamefont {Tiberkevich}, \citenamefont {Xiao},
  \citenamefont {Slavin},\ and\ \citenamefont {Krivorotov}}]{Che17nal}%
  \BibitemOpen
  \bibfield  {author} {\bibinfo {author} {\bibfnamefont {Y.-J.}\ \bibnamefont
  {Chen}}, \bibinfo {author} {\bibfnamefont {H.~K.}\ \bibnamefont {Lee}},
  \bibinfo {author} {\bibfnamefont {R.}~\bibnamefont {Verba}}, \bibinfo
  {author} {\bibfnamefont {J.~A.}\ \bibnamefont {Katine}}, \bibinfo {author}
  {\bibfnamefont {I.}~\bibnamefont {Barsukov}}, \bibinfo {author}
  {\bibfnamefont {V.}~\bibnamefont {Tiberkevich}}, \bibinfo {author}
  {\bibfnamefont {J.~Q.}\ \bibnamefont {Xiao}}, \bibinfo {author}
  {\bibfnamefont {A.~N.}\ \bibnamefont {Slavin}}, \ and\ \bibinfo {author}
  {\bibfnamefont {I.~N.}\ \bibnamefont {Krivorotov}},\ }\href {\doibase
  10.1021/acs.nanolett.6b04725} {\bibfield  {journal} {\bibinfo  {journal}
  {Nano Lett.}\ }\textbf {\bibinfo {volume} {17}},\ \bibinfo {pages} {572}
  (\bibinfo {year} {2017})}\BibitemShut {NoStop}%
\bibitem [{\citenamefont {Merbouche}\ \emph {et~al.}(2021)\citenamefont
  {Merbouche}, \citenamefont {Boventer}, \citenamefont {Haspot}, \citenamefont
  {Fusil}, \citenamefont {Garcia}, \citenamefont {Gou{\'e}r{\'e}},
  \citenamefont {Carr{\'e}t{\'e}ro}, \citenamefont {Vecchiola}, \citenamefont
  {Lebrun}, \citenamefont {Bortolotti}, \citenamefont {Vila}, \citenamefont
  {Bibes}, \citenamefont {Barth{\'e}l{\'e}my},\ and\ \citenamefont
  {Anane}}]{Mer21acs}%
  \BibitemOpen
  \bibfield  {author} {\bibinfo {author} {\bibfnamefont {H.}~\bibnamefont
  {Merbouche}}, \bibinfo {author} {\bibfnamefont {I.}~\bibnamefont {Boventer}},
  \bibinfo {author} {\bibfnamefont {V.}~\bibnamefont {Haspot}}, \bibinfo
  {author} {\bibfnamefont {S.}~\bibnamefont {Fusil}}, \bibinfo {author}
  {\bibfnamefont {V.}~\bibnamefont {Garcia}}, \bibinfo {author} {\bibfnamefont
  {D.}~\bibnamefont {Gou{\'e}r{\'e}}}, \bibinfo {author} {\bibfnamefont
  {C.}~\bibnamefont {Carr{\'e}t{\'e}ro}}, \bibinfo {author} {\bibfnamefont
  {A.}~\bibnamefont {Vecchiola}}, \bibinfo {author} {\bibfnamefont
  {R.}~\bibnamefont {Lebrun}}, \bibinfo {author} {\bibfnamefont
  {P.}~\bibnamefont {Bortolotti}}, \bibinfo {author} {\bibfnamefont
  {L.}~\bibnamefont {Vila}}, \bibinfo {author} {\bibfnamefont {M.}~\bibnamefont
  {Bibes}}, \bibinfo {author} {\bibfnamefont {A.}~\bibnamefont
  {Barth{\'e}l{\'e}my}}, \ and\ \bibinfo {author} {\bibfnamefont
  {A.}~\bibnamefont {Anane}},\ }\href {\doibase 10.1021/acsnano.1c00499}
  {\bibfield  {journal} {\bibinfo  {journal} {ACS Nano}\ }\textbf {\bibinfo
  {volume} {15}},\ \bibinfo {pages} {9775} (\bibinfo {year}
  {2021})}\BibitemShut {NoStop}%
\bibitem [{\citenamefont {Golovchanskiy}\ \emph {et~al.}(2018)\citenamefont
  {Golovchanskiy}, \citenamefont {Abramov}, \citenamefont {Stolyarov},
  \citenamefont {Bolginov}, \citenamefont {Ryazanov}, \citenamefont {Golubov},\
  and\ \citenamefont {Ustinov}}]{Gol18afm}%
  \BibitemOpen
  \bibfield  {author} {\bibinfo {author} {\bibfnamefont {I.~A.}\ \bibnamefont
  {Golovchanskiy}}, \bibinfo {author} {\bibfnamefont {N.~N.}\ \bibnamefont
  {Abramov}}, \bibinfo {author} {\bibfnamefont {V.~S.}\ \bibnamefont
  {Stolyarov}}, \bibinfo {author} {\bibfnamefont {V.~V.}\ \bibnamefont
  {Bolginov}}, \bibinfo {author} {\bibfnamefont {V.~V.}\ \bibnamefont
  {Ryazanov}}, \bibinfo {author} {\bibfnamefont {A.~A.}\ \bibnamefont
  {Golubov}}, \ and\ \bibinfo {author} {\bibfnamefont {A.~V.}\ \bibnamefont
  {Ustinov}},\ }\href {\doibase 10.1002/adfm.201802375} {\bibfield  {journal}
  {\bibinfo  {journal} {Adv. Func. Mater.}\ }\textbf {\bibinfo {volume} {28}},\
  \bibinfo {pages} {1802375} (\bibinfo {year} {2018})}\BibitemShut {NoStop}%
\bibitem [{\citenamefont {Dobrovolskiy}\ \emph
  {et~al.}(2019{\natexlab{b}})\citenamefont {Dobrovolskiy}, \citenamefont
  {Sachser}, \citenamefont {Br{\"a}cher}, \citenamefont {B{\"o}ttcher},
  \citenamefont {Kruglyak}, \citenamefont {Vovk}, \citenamefont {Shklovskij},
  \citenamefont {Huth}, \citenamefont {Hillebrands},\ and\ \citenamefont
  {Chumak}}]{Dob19nph}%
  \BibitemOpen
  \bibfield  {author} {\bibinfo {author} {\bibfnamefont {O.~V.}\ \bibnamefont
  {Dobrovolskiy}}, \bibinfo {author} {\bibfnamefont {R.}~\bibnamefont
  {Sachser}}, \bibinfo {author} {\bibfnamefont {T.}~\bibnamefont
  {Br{\"a}cher}}, \bibinfo {author} {\bibfnamefont {T.}~\bibnamefont
  {B{\"o}ttcher}}, \bibinfo {author} {\bibfnamefont {V.~V.}\ \bibnamefont
  {Kruglyak}}, \bibinfo {author} {\bibfnamefont {R.~V.}\ \bibnamefont {Vovk}},
  \bibinfo {author} {\bibfnamefont {V.~A.}\ \bibnamefont {Shklovskij}},
  \bibinfo {author} {\bibfnamefont {M.}~\bibnamefont {Huth}}, \bibinfo {author}
  {\bibfnamefont {B.}~\bibnamefont {Hillebrands}}, \ and\ \bibinfo {author}
  {\bibfnamefont {A.~V.}\ \bibnamefont {Chumak}},\ }\href {\doibase
  10.1038/s41567-019-0428-5} {\bibfield  {journal} {\bibinfo  {journal} {Nat.
  Phys.}\ }\textbf {\bibinfo {volume} {15}},\ \bibinfo {pages} {477} (\bibinfo
  {year} {2019}{\natexlab{b}})}\BibitemShut {NoStop}%
\bibitem [{\citenamefont {Brandt}(1995)}]{Bra95rpp}%
  \BibitemOpen
  \bibfield  {author} {\bibinfo {author} {\bibfnamefont {E.~H.}\ \bibnamefont
  {Brandt}},\ }\href {http://stacks.iop.org/0034-4885/58/i=11/a=003} {\bibfield
   {journal} {\bibinfo  {journal} {Rep. Progr. Phys.}\ }\textbf {\bibinfo
  {volume} {58}},\ \bibinfo {pages} {1465} (\bibinfo {year}
  {1995})}\BibitemShut {NoStop}%
\bibitem [{\citenamefont {Abrikosov}(2004)}]{Abr04rmp}%
  \BibitemOpen
  \bibfield  {author} {\bibinfo {author} {\bibfnamefont {A.~A.}\ \bibnamefont
  {Abrikosov}},\ }\href {\doibase 10.1103/RevModPhys.76.975} {\bibfield
  {journal} {\bibinfo  {journal} {Rev. Mod. Phys.}\ }\textbf {\bibinfo {volume}
  {76}},\ \bibinfo {pages} {975} (\bibinfo {year} {2004})}\BibitemShut
  {NoStop}%
\bibitem [{\citenamefont {Chumak}\ \emph {et~al.}(2017)\citenamefont {Chumak},
  \citenamefont {Serga},\ and\ \citenamefont {Hillebrands}}]{Chu17jpd}%
  \BibitemOpen
  \bibfield  {author} {\bibinfo {author} {\bibfnamefont {A.~V.}\ \bibnamefont
  {Chumak}}, \bibinfo {author} {\bibfnamefont {A.~A.}\ \bibnamefont {Serga}}, \
  and\ \bibinfo {author} {\bibfnamefont {B.}~\bibnamefont {Hillebrands}},\
  }\href {http://stacks.iop.org/0022-3727/50/i=24/a=244001} {\bibfield
  {journal} {\bibinfo  {journal} {J. Phys. D: Appl. Phys.}\ }\textbf {\bibinfo
  {volume} {50}},\ \bibinfo {pages} {244001} (\bibinfo {year}
  {2017})}\BibitemShut {NoStop}%
\bibitem [{\citenamefont {Dobrovolskiy}\ \emph
  {et~al.}(2020{\natexlab{a}})\citenamefont {Dobrovolskiy}, \citenamefont
  {Vodolazov}, \citenamefont {Porrati}, \citenamefont {Sachser}, \citenamefont
  {Bevz}, \citenamefont {Mikhailov}, \citenamefont {Chumak},\ and\
  \citenamefont {Huth}}]{Dob20nac}%
  \BibitemOpen
  \bibfield  {author} {\bibinfo {author} {\bibfnamefont {O.~V.}\ \bibnamefont
  {Dobrovolskiy}}, \bibinfo {author} {\bibfnamefont {D.~Y.}\ \bibnamefont
  {Vodolazov}}, \bibinfo {author} {\bibfnamefont {F.}~\bibnamefont {Porrati}},
  \bibinfo {author} {\bibfnamefont {R.}~\bibnamefont {Sachser}}, \bibinfo
  {author} {\bibfnamefont {V.~M.}\ \bibnamefont {Bevz}}, \bibinfo {author}
  {\bibfnamefont {M.~Y.}\ \bibnamefont {Mikhailov}}, \bibinfo {author}
  {\bibfnamefont {A.~V.}\ \bibnamefont {Chumak}}, \ and\ \bibinfo {author}
  {\bibfnamefont {M.}~\bibnamefont {Huth}},\ }\href {\doibase
  10.1038/s41467-020-16987-y} {\bibfield  {journal} {\bibinfo  {journal} {Nat.
  Commun.}\ }\textbf {\bibinfo {volume} {11}},\ \bibinfo {pages} {3291}
  (\bibinfo {year} {2020}{\natexlab{a}})}\BibitemShut {NoStop}%
\bibitem [{\citenamefont {Eshbach}\ and\ \citenamefont
  {Damon}(1960)}]{Esh60prv}%
  \BibitemOpen
  \bibfield  {author} {\bibinfo {author} {\bibfnamefont {J.~R.}\ \bibnamefont
  {Eshbach}}\ and\ \bibinfo {author} {\bibfnamefont {R.~W.}\ \bibnamefont
  {Damon}},\ }\href {\doibase 10.1103/PhysRev.118.1208} {\bibfield  {journal}
  {\bibinfo  {journal} {Phys. Rev.}\ }\textbf {\bibinfo {volume} {118}},\
  \bibinfo {pages} {1208} (\bibinfo {year} {1960})}\BibitemShut {NoStop}%
\bibitem [{\citenamefont {Mohseni}\ \emph {et~al.}(2020)\citenamefont
  {Mohseni}, \citenamefont {Hillebrands}, \citenamefont {Pirro},\ and\
  \citenamefont {Kostylev}}]{Moh20prb}%
  \BibitemOpen
  \bibfield  {author} {\bibinfo {author} {\bibfnamefont {M.}~\bibnamefont
  {Mohseni}}, \bibinfo {author} {\bibfnamefont {B.}~\bibnamefont
  {Hillebrands}}, \bibinfo {author} {\bibfnamefont {P.}~\bibnamefont {Pirro}},
  \ and\ \bibinfo {author} {\bibfnamefont {M.}~\bibnamefont {Kostylev}},\
  }\href {\doibase 10.1103/PhysRevB.102.014445} {\bibfield  {journal} {\bibinfo
   {journal} {Phys. Rev. B}\ }\textbf {\bibinfo {volume} {102}},\ \bibinfo
  {pages} {014445} (\bibinfo {year} {2020})}\BibitemShut {NoStop}%
\bibitem [{\citenamefont {Ot\'alora}\ \emph {et~al.}(2016)\citenamefont
  {Ot\'alora}, \citenamefont {Yan}, \citenamefont {Schultheiss}, \citenamefont
  {Hertel},\ and\ \citenamefont {K\'akay}}]{Ota16prl}%
  \BibitemOpen
  \bibfield  {author} {\bibinfo {author} {\bibfnamefont {J.~A.}\ \bibnamefont
  {Ot\'alora}}, \bibinfo {author} {\bibfnamefont {M.}~\bibnamefont {Yan}},
  \bibinfo {author} {\bibfnamefont {H.}~\bibnamefont {Schultheiss}}, \bibinfo
  {author} {\bibfnamefont {R.}~\bibnamefont {Hertel}}, \ and\ \bibinfo {author}
  {\bibfnamefont {A.}~\bibnamefont {K\'akay}},\ }\href {\doibase
  10.1103/PhysRevLett.117.227203} {\bibfield  {journal} {\bibinfo  {journal}
  {Phys. Rev. Lett.}\ }\textbf {\bibinfo {volume} {117}},\ \bibinfo {pages}
  {227203} (\bibinfo {year} {2016})}\BibitemShut {NoStop}%
\bibitem [{\citenamefont {Heinz}\ \emph {et~al.}(2021)\citenamefont {Heinz},
  \citenamefont {Wang}, \citenamefont {Schneider}, \citenamefont {Wei{\ss}},
  \citenamefont {Lentfert}, \citenamefont {L{\"a}gel}, \citenamefont
  {Br{\"a}cher}, \citenamefont {Dubs}, \citenamefont {Dobrovolskiy},
  \citenamefont {Pirro},\ and\ \citenamefont {Chumak}}]{Hei21apl}%
  \BibitemOpen
  \bibfield  {author} {\bibinfo {author} {\bibfnamefont {B.}~\bibnamefont
  {Heinz}}, \bibinfo {author} {\bibfnamefont {Q.}~\bibnamefont {Wang}},
  \bibinfo {author} {\bibfnamefont {M.}~\bibnamefont {Schneider}}, \bibinfo
  {author} {\bibfnamefont {E.}~\bibnamefont {Wei{\ss}}}, \bibinfo {author}
  {\bibfnamefont {A.}~\bibnamefont {Lentfert}}, \bibinfo {author}
  {\bibfnamefont {B.}~\bibnamefont {L{\"a}gel}}, \bibinfo {author}
  {\bibfnamefont {T.}~\bibnamefont {Br{\"a}cher}}, \bibinfo {author}
  {\bibfnamefont {C.}~\bibnamefont {Dubs}}, \bibinfo {author} {\bibfnamefont
  {O.~V.}\ \bibnamefont {Dobrovolskiy}}, \bibinfo {author} {\bibfnamefont
  {P.}~\bibnamefont {Pirro}}, \ and\ \bibinfo {author} {\bibfnamefont {A.~V.}\
  \bibnamefont {Chumak}},\ }\href {\doibase 10.1063/5.0045570} {\bibfield
  {journal} {\bibinfo  {journal} {Appl. Phys. Lett.}\ }\textbf {\bibinfo
  {volume} {118}},\ \bibinfo {pages} {132406} (\bibinfo {year}
  {2021})}\BibitemShut {NoStop}%
\bibitem [{\citenamefont {Gladii}\ \emph {et~al.}(2016)\citenamefont {Gladii},
  \citenamefont {Haidar}, \citenamefont {Henry}, \citenamefont {Kostylev},\
  and\ \citenamefont {Bailleul}}]{Gla16prb}%
  \BibitemOpen
  \bibfield  {author} {\bibinfo {author} {\bibfnamefont {O.}~\bibnamefont
  {Gladii}}, \bibinfo {author} {\bibfnamefont {M.}~\bibnamefont {Haidar}},
  \bibinfo {author} {\bibfnamefont {Y.}~\bibnamefont {Henry}}, \bibinfo
  {author} {\bibfnamefont {M.}~\bibnamefont {Kostylev}}, \ and\ \bibinfo
  {author} {\bibfnamefont {M.}~\bibnamefont {Bailleul}},\ }\href {\doibase
  10.1103/PhysRevB.93.054430} {\bibfield  {journal} {\bibinfo  {journal} {Phys.
  Rev. B}\ }\textbf {\bibinfo {volume} {93}},\ \bibinfo {pages} {054430}
  (\bibinfo {year} {2016})}\BibitemShut {NoStop}%
\bibitem [{\citenamefont {Di}\ \emph {et~al.}(2015)\citenamefont {Di},
  \citenamefont {Zhang}, \citenamefont {Lim}, \citenamefont {Ng}, \citenamefont
  {Kuok}, \citenamefont {Qiu},\ and\ \citenamefont {Yang}}]{Dik15apl}%
  \BibitemOpen
  \bibfield  {author} {\bibinfo {author} {\bibfnamefont {K.}~\bibnamefont
  {Di}}, \bibinfo {author} {\bibfnamefont {V.~L.}\ \bibnamefont {Zhang}},
  \bibinfo {author} {\bibfnamefont {H.~S.}\ \bibnamefont {Lim}}, \bibinfo
  {author} {\bibfnamefont {S.~C.}\ \bibnamefont {Ng}}, \bibinfo {author}
  {\bibfnamefont {M.~H.}\ \bibnamefont {Kuok}}, \bibinfo {author}
  {\bibfnamefont {X.}~\bibnamefont {Qiu}}, \ and\ \bibinfo {author}
  {\bibfnamefont {H.}~\bibnamefont {Yang}},\ }\href {\doibase
  10.1063/1.4907173} {\bibfield  {journal} {\bibinfo  {journal} {Appl. Phys.
  Lett.}\ }\textbf {\bibinfo {volume} {106}},\ \bibinfo {pages} {052403}
  (\bibinfo {year} {2015})}\BibitemShut {NoStop}%
\bibitem [{\citenamefont {B\"ottcher}\ \emph {et~al.}(2021)\citenamefont
  {B\"ottcher}, \citenamefont {Lee}, \citenamefont {Heussner}, \citenamefont
  {Jaiswal}, \citenamefont {Jakob}, \citenamefont {Kl\"aui}, \citenamefont
  {Hillebrands}, \citenamefont {Br\"acher},\ and\ \citenamefont
  {Pirro}}]{Bot21tom}%
  \BibitemOpen
  \bibfield  {author} {\bibinfo {author} {\bibfnamefont {T.}~\bibnamefont
  {B\"ottcher}}, \bibinfo {author} {\bibfnamefont {K.}~\bibnamefont {Lee}},
  \bibinfo {author} {\bibfnamefont {F.}~\bibnamefont {Heussner}}, \bibinfo
  {author} {\bibfnamefont {S.}~\bibnamefont {Jaiswal}}, \bibinfo {author}
  {\bibfnamefont {G.}~\bibnamefont {Jakob}}, \bibinfo {author} {\bibfnamefont
  {M.}~\bibnamefont {Kl\"aui}}, \bibinfo {author} {\bibfnamefont
  {B.}~\bibnamefont {Hillebrands}}, \bibinfo {author} {\bibfnamefont
  {T.}~\bibnamefont {Br\"acher}}, \ and\ \bibinfo {author} {\bibfnamefont
  {P.}~\bibnamefont {Pirro}},\ }\href {\doibase 10.1109/TMAG.2021.3079259}
  {\bibfield  {journal} {\bibinfo  {journal} {IEEE Trans. Magnet.}\ }\textbf
  {\bibinfo {volume} {57}},\ \bibinfo {pages} {1} (\bibinfo {year}
  {2021})}\BibitemShut {NoStop}%
\bibitem [{\citenamefont {Stancil}\ \emph {et~al.}(2006)\citenamefont
  {Stancil}, \citenamefont {Henty}, \citenamefont {Cepni},\ and\ \citenamefont
  {Vant~Hof}}]{Sta06prb}%
  \BibitemOpen
  \bibfield  {author} {\bibinfo {author} {\bibfnamefont {D.~D.}\ \bibnamefont
  {Stancil}}, \bibinfo {author} {\bibfnamefont {B.~E.}\ \bibnamefont {Henty}},
  \bibinfo {author} {\bibfnamefont {A.~G.}\ \bibnamefont {Cepni}}, \ and\
  \bibinfo {author} {\bibfnamefont {J.~P.}\ \bibnamefont {Vant~Hof}},\ }\href
  {\doibase 10.1103/PhysRevB.74.060404} {\bibfield  {journal} {\bibinfo
  {journal} {Phys. Rev. B}\ }\textbf {\bibinfo {volume} {74}},\ \bibinfo
  {pages} {060404} (\bibinfo {year} {2006})}\BibitemShut {NoStop}%
\bibitem [{\citenamefont {Chumak}\ \emph {et~al.}(2010)\citenamefont {Chumak},
  \citenamefont {Dhagat}, \citenamefont {Jander}, \citenamefont {Serga},\ and\
  \citenamefont {Hillebrands}}]{Chu10prb}%
  \BibitemOpen
  \bibfield  {author} {\bibinfo {author} {\bibfnamefont {A.~V.}\ \bibnamefont
  {Chumak}}, \bibinfo {author} {\bibfnamefont {P.}~\bibnamefont {Dhagat}},
  \bibinfo {author} {\bibfnamefont {A.}~\bibnamefont {Jander}}, \bibinfo
  {author} {\bibfnamefont {A.~A.}\ \bibnamefont {Serga}}, \ and\ \bibinfo
  {author} {\bibfnamefont {B.}~\bibnamefont {Hillebrands}},\ }\href {\doibase
  10.1103/PhysRevB.81.140404} {\bibfield  {journal} {\bibinfo  {journal} {Phys.
  Rev. B}\ }\textbf {\bibinfo {volume} {81}},\ \bibinfo {pages} {140404}
  (\bibinfo {year} {2010})}\BibitemShut {NoStop}%
\bibitem [{\citenamefont {Vlaminck}\ and\ \citenamefont
  {Bailleul}(2008)}]{Vla08sci}%
  \BibitemOpen
  \bibfield  {author} {\bibinfo {author} {\bibfnamefont {V.}~\bibnamefont
  {Vlaminck}}\ and\ \bibinfo {author} {\bibfnamefont {M.}~\bibnamefont
  {Bailleul}},\ }\href {\doibase 10.1126/science.1162843} {\bibfield  {journal}
  {\bibinfo  {journal} {Science}\ }\textbf {\bibinfo {volume} {322}},\ \bibinfo
  {pages} {410} (\bibinfo {year} {2008})}\BibitemShut {NoStop}%
\bibitem [{\citenamefont {Dobrovolskiy}(2017)}]{Dob17pcs}%
  \BibitemOpen
  \bibfield  {author} {\bibinfo {author} {\bibfnamefont {O.~V.}\ \bibnamefont
  {Dobrovolskiy}},\ }\href {\doibase
  http://doi.org/10.1016/j.physc.2016.07.008} {\bibfield  {journal} {\bibinfo
  {journal} {Physica C}\ }\textbf {\bibinfo {volume} {533}},\ \bibinfo {pages}
  {80} (\bibinfo {year} {2017})}\BibitemShut {NoStop}%
\bibitem [{\citenamefont {Plourde}(2009)}]{Plo09tas}%
  \BibitemOpen
  \bibfield  {author} {\bibinfo {author} {\bibfnamefont {B.~L.~T.}\
  \bibnamefont {Plourde}},\ }\href {\doibase doi:10.1109/TASC.2009.2028873}
  {\bibfield  {journal} {\bibinfo  {journal} {IEEE Trans. Appl. Supercond.}\
  }\textbf {\bibinfo {volume} {19}},\ \bibinfo {pages} {3698} (\bibinfo {year}
  {2009})}\BibitemShut {NoStop}%
\bibitem [{\citenamefont {Shklovskij}\ \emph {et~al.}(2014)\citenamefont
  {Shklovskij}, \citenamefont {Sosedkin},\ and\ \citenamefont
  {Dobrovolskiy}}]{Shk14pcm}%
  \BibitemOpen
  \bibfield  {author} {\bibinfo {author} {\bibfnamefont {V.~A.}\ \bibnamefont
  {Shklovskij}}, \bibinfo {author} {\bibfnamefont {V.~V.}\ \bibnamefont
  {Sosedkin}}, \ and\ \bibinfo {author} {\bibfnamefont {O.~V.}\ \bibnamefont
  {Dobrovolskiy}},\ }\href {http://stacks.iop.org/0953-8984/26/i=2/a=025703}
  {\bibfield  {journal} {\bibinfo  {journal} {J. Phys.: Cond. Matt.}\ }\textbf
  {\bibinfo {volume} {26}},\ \bibinfo {pages} {025703} (\bibinfo {year}
  {2014})}\BibitemShut {NoStop}%
\bibitem [{\citenamefont {Dobrovolskiy}\ \emph {et~al.}(2015)\citenamefont
  {Dobrovolskiy}, \citenamefont {Huth},\ and\ \citenamefont
  {Shklovskij}}]{Dob15met}%
  \BibitemOpen
  \bibfield  {author} {\bibinfo {author} {\bibfnamefont {O.~V.}\ \bibnamefont
  {Dobrovolskiy}}, \bibinfo {author} {\bibfnamefont {M.}~\bibnamefont {Huth}},
  \ and\ \bibinfo {author} {\bibfnamefont {V.~A.}\ \bibnamefont {Shklovskij}},\
  }\href {\doibase http://dx.doi.org/10.1063/1.4934487} {\bibfield  {journal}
  {\bibinfo  {journal} {Appl. Phys. Lett.}\ }\textbf {\bibinfo {volume}
  {107}},\ \bibinfo {pages} {162603} (\bibinfo {year} {2015})}\BibitemShut
  {NoStop}%
\bibitem [{\citenamefont {Dobrovolskiy}\ \emph
  {et~al.}(2020{\natexlab{b}})\citenamefont {Dobrovolskiy}, \citenamefont
  {Begun}, \citenamefont {Bevz}, \citenamefont {Sachser},\ and\ \citenamefont
  {Huth}}]{Dob20pra}%
  \BibitemOpen
  \bibfield  {author} {\bibinfo {author} {\bibfnamefont {O.}~\bibnamefont
  {Dobrovolskiy}}, \bibinfo {author} {\bibfnamefont {E.}~\bibnamefont {Begun}},
  \bibinfo {author} {\bibfnamefont {V.}~\bibnamefont {Bevz}}, \bibinfo {author}
  {\bibfnamefont {R.}~\bibnamefont {Sachser}}, \ and\ \bibinfo {author}
  {\bibfnamefont {M.}~\bibnamefont {Huth}},\ }\href {\doibase
  10.1103/PhysRevApplied.13.024012} {\bibfield  {journal} {\bibinfo  {journal}
  {Phys. Rev. Appl.}\ }\textbf {\bibinfo {volume} {13}},\ \bibinfo {pages}
  {024012} (\bibinfo {year} {2020}{\natexlab{b}})}\BibitemShut {NoStop}%
\bibitem [{\citenamefont {W\"ordenweber}\ \emph {et~al.}(2012)\citenamefont
  {W\"ordenweber}, \citenamefont {Hollmann}, \citenamefont {Schubert},
  \citenamefont {Kutzner},\ and\ \citenamefont {Panaitov}}]{Wor12prb}%
  \BibitemOpen
  \bibfield  {author} {\bibinfo {author} {\bibfnamefont {R.}~\bibnamefont
  {W\"ordenweber}}, \bibinfo {author} {\bibfnamefont {E.}~\bibnamefont
  {Hollmann}}, \bibinfo {author} {\bibfnamefont {J.}~\bibnamefont {Schubert}},
  \bibinfo {author} {\bibfnamefont {R.}~\bibnamefont {Kutzner}}, \ and\
  \bibinfo {author} {\bibfnamefont {G.}~\bibnamefont {Panaitov}},\ }\href
  {\doibase 10.1103/PhysRevB.85.064503} {\bibfield  {journal} {\bibinfo
  {journal} {Phys. Rev. B}\ }\textbf {\bibinfo {volume} {85}},\ \bibinfo
  {pages} {064503} (\bibinfo {year} {2012})}\BibitemShut {NoStop}%
\bibitem [{\citenamefont {Dobrovolskiy}\ and\ \citenamefont
  {Huth}(2015)}]{Dob15apl}%
  \BibitemOpen
  \bibfield  {author} {\bibinfo {author} {\bibfnamefont {O.~V.}\ \bibnamefont
  {Dobrovolskiy}}\ and\ \bibinfo {author} {\bibfnamefont {M.}~\bibnamefont
  {Huth}},\ }\href {\doibase http://dx.doi.org/10.1063/1.4917229} {\bibfield
  {journal} {\bibinfo  {journal} {Appl. Phys. Lett.}\ }\textbf {\bibinfo
  {volume} {106}},\ \bibinfo {pages} {142601} (\bibinfo {year}
  {2015})}\BibitemShut {NoStop}%
\bibitem [{\citenamefont {Buzdin}(2005)}]{Buz05rmp}%
  \BibitemOpen
  \bibfield  {author} {\bibinfo {author} {\bibfnamefont {A.~I.}\ \bibnamefont
  {Buzdin}},\ }\href {\doibase 10.1103/RevModPhys.77.935} {\bibfield  {journal}
  {\bibinfo  {journal} {Rev. Mod. Phys.}\ }\textbf {\bibinfo {volume} {77}},\
  \bibinfo {pages} {935} (\bibinfo {year} {2005})}\BibitemShut {NoStop}%
\bibitem [{\citenamefont {Kompaniiets}\ \emph {et~al.}(2014)\citenamefont
  {Kompaniiets}, \citenamefont {Dobrovolskiy}, \citenamefont {Neetzel},
  \citenamefont {Porrati}, \citenamefont {Br\"otz}, \citenamefont {Ensinger},\
  and\ \citenamefont {Huth}}]{Kom14apl}%
  \BibitemOpen
  \bibfield  {author} {\bibinfo {author} {\bibfnamefont {M.}~\bibnamefont
  {Kompaniiets}}, \bibinfo {author} {\bibfnamefont {O.~V.}\ \bibnamefont
  {Dobrovolskiy}}, \bibinfo {author} {\bibfnamefont {C.}~\bibnamefont
  {Neetzel}}, \bibinfo {author} {\bibfnamefont {F.}~\bibnamefont {Porrati}},
  \bibinfo {author} {\bibfnamefont {J.}~\bibnamefont {Br\"otz}}, \bibinfo
  {author} {\bibfnamefont {W.}~\bibnamefont {Ensinger}}, \ and\ \bibinfo
  {author} {\bibfnamefont {M.}~\bibnamefont {Huth}},\ }\href {\doibase
  10.1063/1.4863980} {\bibfield  {journal} {\bibinfo  {journal} {Appl. Phys.
  Lett.}\ }\textbf {\bibinfo {volume} {104}},\ \bibinfo {pages} {052603}
  (\bibinfo {year} {2014})}\BibitemShut {NoStop}%
\bibitem [{\citenamefont {Dobrovolskiy}\ \emph {et~al.}(2012)\citenamefont
  {Dobrovolskiy}, \citenamefont {Begun}, \citenamefont {Huth},\ and\
  \citenamefont {Shklovskij}}]{Dob12njp}%
  \BibitemOpen
  \bibfield  {author} {\bibinfo {author} {\bibfnamefont {O.~V.}\ \bibnamefont
  {Dobrovolskiy}}, \bibinfo {author} {\bibfnamefont {E.}~\bibnamefont {Begun}},
  \bibinfo {author} {\bibfnamefont {M.}~\bibnamefont {Huth}}, \ and\ \bibinfo
  {author} {\bibfnamefont {V.~A.}\ \bibnamefont {Shklovskij}},\ }\href
  {http://stacks.iop.org/1367-2630/14/i=11/a=113027} {\bibfield  {journal}
  {\bibinfo  {journal} {New J. Phys.}\ }\textbf {\bibinfo {volume} {14}},\
  \bibinfo {pages} {113027} (\bibinfo {year} {2012})}\BibitemShut {NoStop}%
\bibitem [{\citenamefont {Gubin}\ \emph {et~al.}(2005)\citenamefont {Gubin},
  \citenamefont {Il'in}, \citenamefont {Vitusevich}, \citenamefont {Siegel},\
  and\ \citenamefont {Klein}}]{Gub05prb}%
  \BibitemOpen
  \bibfield  {author} {\bibinfo {author} {\bibfnamefont {A.~I.}\ \bibnamefont
  {Gubin}}, \bibinfo {author} {\bibfnamefont {K.~S.}\ \bibnamefont {Il'in}},
  \bibinfo {author} {\bibfnamefont {S.~A.}\ \bibnamefont {Vitusevich}},
  \bibinfo {author} {\bibfnamefont {M.}~\bibnamefont {Siegel}}, \ and\ \bibinfo
  {author} {\bibfnamefont {N.}~\bibnamefont {Klein}},\ }\href {\doibase
  10.1103/PhysRevB.72.064503} {\bibfield  {journal} {\bibinfo  {journal} {Phys.
  Rev. B}\ }\textbf {\bibinfo {volume} {72}},\ \bibinfo {pages} {064503}
  (\bibinfo {year} {2005})}\BibitemShut {NoStop}%
\bibitem [{\citenamefont {Kim}\ \emph {et~al.}(2003)\citenamefont {Kim},
  \citenamefont {Kim}, \citenamefont {Hong}, \citenamefont {Hwang},\ and\
  \citenamefont {Hahn}}]{Kim03cry}%
  \BibitemOpen
  \bibfield  {author} {\bibinfo {author} {\bibfnamefont {D.~H.}\ \bibnamefont
  {Kim}}, \bibinfo {author} {\bibfnamefont {K.~T.}\ \bibnamefont {Kim}},
  \bibinfo {author} {\bibfnamefont {H.~G.}\ \bibnamefont {Hong}}, \bibinfo
  {author} {\bibfnamefont {J.~S.}\ \bibnamefont {Hwang}}, \ and\ \bibinfo
  {author} {\bibfnamefont {T.~S.}\ \bibnamefont {Hahn}},\ }\href {\doibase
  10.1016/S0011-2275(03)00163-2} {\bibfield  {journal} {\bibinfo  {journal}
  {Cryogenics}\ }\textbf {\bibinfo {volume} {43}},\ \bibinfo {pages} {561}
  (\bibinfo {year} {2003})}\BibitemShut {NoStop}%
\bibitem [{\citenamefont {Dobrovolskiy}\ and\ \citenamefont
  {Huth}(2012)}]{Dob12tsf}%
  \BibitemOpen
  \bibfield  {author} {\bibinfo {author} {\bibfnamefont {O.~V.}\ \bibnamefont
  {Dobrovolskiy}}\ and\ \bibinfo {author} {\bibfnamefont {M.}~\bibnamefont
  {Huth}},\ }\href {\doibase 10.1016/j.tsf.2012.04.083} {\bibfield  {journal}
  {\bibinfo  {journal} {Thin Solid Films}\ }\textbf {\bibinfo {volume} {520}},\
  \bibinfo {pages} {5985} (\bibinfo {year} {2012})}\BibitemShut {NoStop}%
\bibitem [{\citenamefont {Krawczyk}\ and\ \citenamefont
  {Grundler}(2014)}]{Kra14pcm}%
  \BibitemOpen
  \bibfield  {author} {\bibinfo {author} {\bibfnamefont {M.}~\bibnamefont
  {Krawczyk}}\ and\ \bibinfo {author} {\bibfnamefont {D.}~\bibnamefont
  {Grundler}},\ }\href {http://stacks.iop.org/0953-8984/26/i=12/a=123202}
  {\bibfield  {journal} {\bibinfo  {journal} {J. Phys.: Condens. Matt.}\
  }\textbf {\bibinfo {volume} {26}},\ \bibinfo {pages} {123202} (\bibinfo
  {year} {2014})}\BibitemShut {NoStop}%
\bibitem [{\citenamefont {Koshelev}\ and\ \citenamefont
  {Vinokur}(1994)}]{Kos94prl}%
  \BibitemOpen
  \bibfield  {author} {\bibinfo {author} {\bibfnamefont {A.~E.}\ \bibnamefont
  {Koshelev}}\ and\ \bibinfo {author} {\bibfnamefont {V.~M.}\ \bibnamefont
  {Vinokur}},\ }\href {\doibase 10.1103/PhysRevLett.73.3580} {\bibfield
  {journal} {\bibinfo  {journal} {Phys. Rev. Lett.}\ }\textbf {\bibinfo
  {volume} {73}},\ \bibinfo {pages} {3580} (\bibinfo {year}
  {1994})}\BibitemShut {NoStop}%
\bibitem [{\citenamefont {Kalinikos}\ and\ \citenamefont
  {Slavin}(1986)}]{Kal86jpc}%
  \BibitemOpen
  \bibfield  {author} {\bibinfo {author} {\bibfnamefont {B.~A.}\ \bibnamefont
  {Kalinikos}}\ and\ \bibinfo {author} {\bibfnamefont {A.~N.}\ \bibnamefont
  {Slavin}},\ }\href {http://stacks.iop.org/0022-3719/19/i=35/a=014} {\bibfield
   {journal} {\bibinfo  {journal} {J. Phys. C}\ }\textbf {\bibinfo {volume}
  {19}},\ \bibinfo {pages} {7013} (\bibinfo {year} {1986})}\BibitemShut
  {NoStop}%
\bibitem [{\citenamefont {Pompeo}\ and\ \citenamefont
  {Silva}(2008)}]{Pom08prb}%
  \BibitemOpen
  \bibfield  {author} {\bibinfo {author} {\bibfnamefont {N.}~\bibnamefont
  {Pompeo}}\ and\ \bibinfo {author} {\bibfnamefont {E.}~\bibnamefont {Silva}},\
  }\href {\doibase 10.1103/PhysRevB.78.094503} {\bibfield  {journal} {\bibinfo
  {journal} {Phys. Rev. B}\ }\textbf {\bibinfo {volume} {78}},\ \bibinfo
  {pages} {094503} (\bibinfo {year} {2008})}\BibitemShut {NoStop}%
\bibitem [{\citenamefont {Dobrovolskiy}\ \emph {et~al.}(2018)\citenamefont
  {Dobrovolskiy}, \citenamefont {Sachser}, \citenamefont {Huth}, \citenamefont
  {Shklovskij}, \citenamefont {Vovk}, \citenamefont {Bevz},\ and\ \citenamefont
  {Tsindlekht}}]{Dob18apl}%
  \BibitemOpen
  \bibfield  {author} {\bibinfo {author} {\bibfnamefont {O.~V.}\ \bibnamefont
  {Dobrovolskiy}}, \bibinfo {author} {\bibfnamefont {R.}~\bibnamefont
  {Sachser}}, \bibinfo {author} {\bibfnamefont {M.}~\bibnamefont {Huth}},
  \bibinfo {author} {\bibfnamefont {V.~A.}\ \bibnamefont {Shklovskij}},
  \bibinfo {author} {\bibfnamefont {R.~V.}\ \bibnamefont {Vovk}}, \bibinfo
  {author} {\bibfnamefont {V.~M.}\ \bibnamefont {Bevz}}, \ and\ \bibinfo
  {author} {\bibfnamefont {M.~I.}\ \bibnamefont {Tsindlekht}},\ }\href
  {\doibase 10.1063/1.5028213} {\bibfield  {journal} {\bibinfo  {journal}
  {Appl. Phys. Lett.}\ }\textbf {\bibinfo {volume} {112}},\ \bibinfo {pages}
  {152601} (\bibinfo {year} {2018})}\BibitemShut {NoStop}%
\end{thebibliography}

%

\end{document}